\documentclass[preprint,12pt]{elsarticle}

\usepackage{amssymb}
\usepackage{amsthm}

\biboptions{}

\def\dbb{$\beta\beta$}

\newcommand\T{\rule{0pt}{2.2ex}}
\newcommand\B{\rule[-1.6ex]{0pt}{0pt}}

\usepackage{lscape}

\journal{Nuclear Physics A}

\begin{document}

\begin{frontmatter}



\title{Resonance enhancement of neutrinoless double electron capture}


\author[label1,label2]{M.I.~Krivoruchenko}
\author[label3,label4]{Fedor \v Simkovic}
\author[label5]{Dieter Frekers}
\author[label6]{Amand Faessler}

\address[label1]{Institute for Theoretical and Experimental Physics$\mathrm{,}$ B. Cheremushkinskaya 25 \\117218 Moscow, Russia}
\address[label2]{Department of Nano-, Bio-, Information and Cognitive Technologies\\
Moscow Institute of Physics and Technology$\mathrm{,}$ 9 Institutskii per. \\
141700 Dolgoprudny, Moscow Region, Russia}
\address[label3]{Bogoliubov Laboratory of Theoretical Physics, JINR \\
141980 Dubna, Moscow Region, Russia} 
\address[label4]{Department of Nuclear Physics
and Biophysics, Comenius University \\
Mlynsk\'a dolina F1, SK-842 48 Bratislava, Slovakia}
\address[label5]{Institut f\"{u}r Kernphysik, Universit\"{a}t  M\"{u}nster$\mathrm{,}$ Wilhelm-Klemm-Str. 9 \\
48149 M\"{u}nster, Germany}
\address[label6]{Institut f\"{u}r Theoretische Physik$\mathrm{,}$ T\"{u}bingen
Universit\"{a}t$\mathrm{,}$ Auf der Morgenstelle 14 \\
D-72076 T\"{u}bingen, Germany}

\begin{abstract}
The process of neutrinoless double electron ($0\nu$ECEC) capture is
revisited for those cases where the two participating atoms are nearly degenerate in mass. 
The theoretical framework is the formalism of an oscillation of two atoms with different total 
lepton number (and parity), one of which can be in an excited state so that mass degeneracy is realized. 
In such a case and assuming light Majorana neutrinos, the two atoms will be in a mixed configuration with 
respect to the weak interaction. A resonant enhancement of transitions between such pairs of atoms will occur, 
which could be detected by the subsequent electromagnetic de-excitation of the excited state of the daughter 
atom and nucleus. Available data of atomic masses, as well as  nuclear and atomic  excitations are used to 
select the most likely candidates for the resonant transitions. Assuming an effective mass for 
the Majorana neutrino of 1 eV, some half-lives are predicted to be as low as $10^{22}$ years in the 
unitary limit. It is argued that, in order to obtain more accurate predictions for the $0\nu$ECEC half-lives, 
precision mass measurements of the atoms involved are necessary, which can readily be accomplished 
by today's high precision Penning traps. Further advancements  also require a better understanding of high-lying excited states of the final nuclei (i.e. excitation energy, angular momentum and parity) and the calculation of the nuclear matrix elements.
\end{abstract}



\end{frontmatter}

\renewcommand{\baselinestretch}{1.0}
\renewcommand{\theequation}{\thesection.\arabic{equation}}

\section{Introduction}
\setcounter{equation}{0}

The question as to whether massive neutrinos obey a Dirac or a Majorana
symmetry, presently constitutes  one of the most important unresolved problems
of particle physics and astrophysics. If  neutrinos are Dirac particles, i.e.
if  neutrino and antineutrino are fundamentally different, then total lepton
number $L=L_e + L_\mu + L_\tau$ must be conserved. Contrary, if neutrinos are
Majorana particles, i.e. if  neutrino and  antineutrino are identical
particles, then  lepton number conservation is not required anymore. Indeed,
lepton number (LN) conservation is one of the most obscure appearances in the
Standard Model of elementary particles, since there is no known
fundamental principle or symmetry, which would require this.

Already in 1939 Furry \cite{FURR39} noticed that the exchange of neutrinos
(later termed Majorana neutrinos) between two neutrons could lead to the
production of two protons and two electrons in the reaction
\begin{eqnarray}
(A, Z) \to (A, Z + 2) + e^- + e^-. \label{DBD}
\end{eqnarray}
Today, such a reaction is termed the neutrinoless double-beta ($0\nu\beta \beta
$) decay, and an observation of this reaction  is still the only unambiguous
way  to identify the Majorana character of the neutrino~\cite{schv}. Over the
years, considerable efforts from experimentalists and theorists alike have been
devoted to this process (for reviews see Ref. \cite{dbdreviews}). Although of
fundamental importance, this process is  unfortunately characterized by an
excessively low rate, which poses a significant challenge to any experiment.

Assuming that the light neutrino mixing mechanism provides the dominant
contribution to the $0\nu\beta \beta $-decay,
the decay rate for a given isotope $(A,Z)$ is simply given by the product of
the effective Majorana neutrino mass
squared $|m_{\beta\beta}|^2$, the known 3-body phase-space factor, and the much
less well-known nuclear matrix element squared,
which is particular to every nuclear transition under study. The phase-space
factor contains a dependence on the nuclear charge
$Z$ ($\sim Z^2$), the Q-value of the reaction ($\sim Q^5$) and the
Fermi-coupling constant ($\sim G_F^4$).
The main objective of every experimental $0\nu\beta\beta$-decay search is the
determination of the absolute value of the effective
Majorana neutrino mass $|m_{\beta\beta}|$. However, a mere observation of the
decay would already constitute a significant
advancement in neutrino physics.

In the 3-neutrino mixing scenario, the effective Majorana neutrino mass
$m_{\beta \beta}$ takes the form
\begin{equation}
m_{\beta\beta}  =  \sum\limits_{i = 1}^{3} {U_{ei}^2 } m_i. \label{e:4}
\end{equation}
Here, $U_{ei}$ ($i=1,2,3$) are the elements of the
Pontecorvo-Maki-Nakagawa-Sakata (PMNS) neutrino mixing matrix, which mixes the
mass eigenstates $m_i$ in weak interaction. It contains the usual 3-neutrino
mixing angles plus a CP-violating phase, which appears in oscillations, and two
additional Majorana phases, $\phi_1,\phi_2$.

The recent claim for an observation of the $0\nu\beta^-\beta^-$-decay of
$^{76}$Ge with $T^{0\nu}_{1/2} = 2.23^{+0.44}_{-0.31}\times 10^{25}$ years
\cite{evidence} implies $|m_{\beta\beta}|\simeq 0.18-0.30$ eV by assuming the
renormalized QRPA (RQRPA) nuclear matrix element and its uncertainty given in
Ref. \cite{src}. The goal of the upcoming GERDA experiment \cite{gerda} is to
put this claim to a test by improving the sensitivity limit of the detection by
more than an order of magnitude. The next generation experiments, which will be
using several other candidate nuclei, will eventually be able to achieve this
goal as well~\cite{avig}.

The value of the effective Majorana mass, as it appears in Eq.~(\ref{e:4}),
contains several dependencies on phases and masses. Because of experimental
uncertainties, different mass scenarios, like  the normal  ($m_3 \gg m_2 \gg
m_1$) or inverted ($m_2 > m_1 \gg m_3$) hierarchy scenario, or degenerate ($m_3
\approx m_2 \approx m_1$) or non-degenerate cases,  can presently still be
entertained, which allow  a wide range of possible mass values for
$|m_{\beta\beta}|$,  even zero in the most extreme and unfortunate situation of
the normal hierarchy scenario \cite{hierar}. Though even in that case the
$0\nu\beta\beta$ will still be allowed due to a contribution from the mass term
in the neutrino propagator~\cite{pepa}, which one usually neglects, its decay
rate would be utterly unobservable.

Recently, there has been an increased  theoretical and experimental interest to
another LN violating process, which is the neutrinoless double electron capture
($0\nu$ECEC) \cite{sujwy,frekers,barab1,barab2}. In this reaction two bound
electrons from the atomic shell are captured by two protons, thereby lowering
the charge of the final nucleus by two units:
\begin{eqnarray}
(A, Z)  + e_b^- + e_b^- \to (A, Z - 2)^{**}. \label{ECEC}
\end{eqnarray}
\noindent Here, the two asterisks denote the possibility of leaving the system
in an  excited nuclear and/or atomic state, the latter being characterized by
two vacancies in the  electron shell of the otherwise neutral atom.
The energy excess given by the Q-value of the reaction must still
be carried away by an extra photon, in order to conserve energy. This is unlike
the 2-neutrino case, where the neutrinos can provide the energy balance. Thus,
the reaction in Eq.~(\ref{ECEC}) could in principle  be detected by monitoring
the $X$-rays or Auger electrons emitted from excited electron shell of the
atom, the electromagnetic decay of the excited nucleus (in case of a
non-ground-state transition) and the extra photon, whose energy would be
\begin{equation}
E_{h\nu}=Q-E_{\rm atom}^x - E_{\rm nucl}^x. \label{xtra-photon}
\end{equation}
We note that $0\nu$ECEC was considered by Winter \cite{WINT55} already in 1955.

The signature of the $0\nu$ECEC process is, therefore,  different from the
signature of the $0\nu\beta\beta$ decay and would also require  rather
different coincident detection techniques. On the other hand, the coupling to
an extra photon and/or X-ray clearly makes the half-life excessively long to
the extent that this process has not been considered a valid experimental
option altogether.

The situation changes, however, if the energy difference in
Eq.~(\ref{xtra-photon}) approaches zero and no extra photon is required.
This has been discussed by Bernab\'eu, De Rujula, and Jarlskog \cite{DERU}, who
pointed  to the possibility of a resonant enhancement of the $0\nu$ECEC decay
in case of a mass degeneracy between the initial and final nucleus.
Their best candidate case was $^{112}$Sn, where the $0\nu$ECEC
double K-shell capture process would lead to an exited $0^+$ state at 1871~keV
in the final nucleus $^{112}$Cd. This possibility was recently excluded by a
new mass difference measurement performed at Jyv\"askyl\"a ($\Delta M  =
1919.82(16)$~ keV)~\cite{RAKH09}, where it was shown that the energy to be
paid by the double K-shell vacancy would not leave enough energy available for
the excitation of the 1871 keV state in  $^{112}$Cd.

The $0\nu$ECEC decays became a subject to a detailed theoretical
treatment by Sujkowski and Wycech \cite{sujwy}, who used a perturbative
approach. Their conclusion was that an exact energy degeneracy could make the
$0\nu$ECEC reaction competitive  to the  $0\nu\beta^-\beta^-$-decay. However, a
case with an exact energy degeneracy could not be identified.

Recently, another case for a near mass-degeneracy was found and discussed in
Ref. \cite{frekers}. Here, it was argued that the 1204~keV state in $^{74}$Ge
would be nearly degenerate to the ground state of the atomic nucleus $^{74}$Se
in case of a double L-capture process and given the experimental errors on the
masses. A new mass difference measurement performed by Kolhinen {\it et
al.}~\cite{kolhinen}  essentially confirmed the previous central mass
difference value of $\Delta M = 1209.169(49)$ keV, however, with much higher
precision. These authors therefore  excluded a complete mass degeneracy with
the 1204 keV state in $^{74}$Ge, even in the case of a double L-capture, where
the atomic  energy for a double L-vacancy would amount to an extra 2.9 keV on
top of the 1204 keV~\cite{frekers}.  Prior to this, two experiments  had
already been  performed by Barabash {\it et al.}~\cite{barab1,barab2}, which
gave lower bounds for the half-lives of $^{74}\mathrm{Se}$ and
$^{112}\mathrm{Sn}$:
\begin{eqnarray*}
T_{1/2}^{0\nu\mathrm{ECEC}} (^{74}\mathrm{Se}) &\geq& 5.5 \times 10^{18}
~\mathrm{y},  \\
T_{1/2}^{0\nu\mathrm{ECEC}}(^{112}\mathrm{Sn}) &\geq& 9.2 \times 10^{19}
~\mathrm{y}.
\end{eqnarray*}

In this paper we present a new theoretical framework for the calculation of resonant 
$0\nu$ECEC transitions, namely the oscillation of atoms.
An improved theoretical description of the process includes the determination
of relevant matrix elements for the most favored cases of capture of the
$s_{1/2}$ and $p_{1/2}$ electrons. The $0\nu$ECEC transitions without and with
the spatial parity violation are considered. Further, we provide an updated
list of the most likely resonant transitions taking new nuclear spectroscopic
data into account and using recent accurate measurements of $Q$ values for
several nuclei~\cite{kolhinen,RAKH09,penning1,penning2,penning3,SCIE09}.  
The selection of transitions is also based on 
accurate treatment of spin-tensor structures that arise in a product of the nuclear matrix elements 
and the electron wave functions of atomic shells. The reverse reaction
\begin{eqnarray}
(A, Z) \to (A, Z + 2)^{**} + e_b^- + e_b^- \label{EPEP}
\end{eqnarray}
of a neutrinoless production of two bound electrons ($0\nu$EPEP) will  also be
discussed.

The outline of the paper is as follows:
First we discuss the mixing of atoms with different lepton charges. This effect 
leads to the oscillations of atoms. In Sect. 2, we discuss the relevant formalism of the
oscillations. We will show that the oscillation of stable atoms
produces a too small effect to be measured experimentally. However, oscillation
between a stable and an excited atom can lead to a resonant enhancement of lepton
number violating decays.

Sect. 3 presents the estimated half-lives of the decays. In the calculations 
we use the data on the Auger and radiative widths of excited electron shells 
and the information on the Coulomb interaction energy of two electron holes.
We consider nuclei with arbitrary spin-parity and take into account the fact that the spin-parity uniquely determines a combination 
of upper and lower components of the relativistic electron wave functions entering the matrix elements associated with the capture.
In Appendix A the procedure of averaging the electron wave functions over the nucleus is discussed. 
The transition matrix elements are derived for the $J^{\pi}_f = 0^{\pm}_f$ and $1^{\pm}_f$ states of the daughter nuclei in Appendix B. 
The problem of calculating matrix elements is very complicated, and the result depends sensitively on the particular transition. 
We identify the most promising nuclei in the search for $0\nu$ECEC decays. Such nuclei will continue to be analyzed in future. 
In this paper, the half-lives are normalized to the nuclear matrix element $\mathcal{M}^{0\nu}(0^{+}_i \to J^{\pi}_f) = 6$, 
which is close to the maximum evaluated value of the matrix elements for medium-heavy nuclei. 
In Sect. 3, we also give a complete list of the most likely resonance transitions, in which the unitary limit of resonant enhancement gives half-lives 
of less than $ 10^{27} $ years for $ | m_ {\beta \beta} | = 1 $ eV.
We argue that accurate measurements
of the mass differences between
initial and final states of the nuclei are necessary, if future experiments of
$0\nu$ECEC decays
with half-lives below $10^{27}$ years were to  become a possibility.
Experimental signatures of $0\nu$ECEC decays are
discussed in Sect. 4.

\section{Lepton number violating transitions between ground state and excited
atoms}
\setcounter{equation}{0}

If lepton number is not conserved, then the weak interaction mixing between a
pair of neutral atoms $(A,Z)$ and $(A,Z \pm 2)$ is a natural occurrence, which
leads to an oscillation between these two many-body quantum systems. In the
present description we focus on a system, in which one of the atoms (usually
the daughter atom) is left in an excited atomic or nuclear state. In fact, for
EC processes the daughter atomic system is always excited, as the capture
process always leaves a vacancy in the electron shell. If the ECEC Q-value is
of the order of the excitation of the atomic shell with two electron vacancies,
one may expect a resonant-like  transition. A few examples do exist in  the
nuclear chart, which have this property. On the other hand, if the   ECEC
Q-value is significantly larger than the atomic excitation, one may find a
situation, where an excited nuclear state matches the available energy (i.e.
$Q-E_{\rm atom}^x = E_{\rm nucl}^x$), allowing again a resonant-like transition
to an excited nuclear state. The latter type of oscillations may even have a
practical experimental signature: one or even several X-ray photons or
Auger electrons from the de-excitation of the atomic shell being coincident
with a $\gamma$-ray (or a cascade of $\gamma$-rays) from the de-excitation of
the nucleus. In fact, the detection of a coincident $\gamma$-ray cascade, if
existent,  may already be sufficient to uniquely identify the transition. It
may be worth re-iterating that any such  transition requires the neutrino to be
of Majorana type, as there is no phase space available for the emission
  two extra neutrinos.

The present description of a resonant enhancement of the ECEC transition will
be done in the context of oscillations. We wish to point out, that our results
are  consistent with the results of Bernab\'eu et al.~\cite{DERU} and Sujkowski
and Wycech~\cite{sujwy} for the physically interesting case, where the
frequency of oscillations is much smaller than the width of the excited atom.
In the opposite limit, when the frequency of oscillations is high, the standard
formulas of the time evolution of a two-level system are retained.

\subsection{Oscillations in arbitrary systems}

Specific features of the oscillations in the system of two atoms were discussed
earlier in Ref.~\cite{SIM08}. Two coupled oscillators, one of which 
experiences friction, constitute the mechanical analogue of the system, which
we are considering.

Lepton number violating interactions induce transitions $(A,Z) \to (A,Z \pm
2)^{**}$. These transitions can be described phenomenologically by  $ 2 \times
2 $ non-Hermitian Hamiltonian matrix
\begin{eqnarray}
H_{\mathrm{eff}} &=& \left(
\begin{array}{cc}
M_i & V \\
V^* & M_f - \frac{i}{2}\Gamma
\end{array}
\right).  \label{HAMI}
\end{eqnarray}
Here $M_i$ and $M_f$ are the masses of the initial and final atom. $\Gamma $ is
the decay width of the excited daughter atom. The off-diagonal matrix elements of
$ H_{\mathrm{eff}}$ are complex conjugate. The transition potential $V$ can
always be made real by changing the phase of one of the states, i.e. $V=V^*$.
The diagonal matrix elements of the Hamiltonian are determined by strong and
electromagnetic interactions, which conserve  lepton number. The off-diagonal
elements provide the mixing of the neutral atoms, and thereby, violate lepton
number by two units as a result of the weak interaction with massive Majorana
neutrinos. Using the Pauli matrices, the Hamiltonian can be written as:
\begin{eqnarray}
H_{\mathrm{eff}} = M_{+} + V \sigma_1 + M_{-}\sigma_3,
\end{eqnarray}
where
\begin{equation}
M_{\pm} = \frac{M_i \pm M_f}{2} \mp \frac{i}{4} \Gamma.
\label{mp_mm}\end{equation} The evolution operator $e^{-i H_{\mathrm{eff}} t}$
can be expanded over the Pauli matrices to give
\begin{eqnarray}
e^{-i H_{\mathrm{eff}} t} = e^{-iM_{+}t} \left( \cos(\Omega t) - i
\frac{V\sigma_1 + M_{-}\sigma_3}{\Omega}\sin(\Omega t) \right), 
\label{EVOLMATR}
\end{eqnarray}
where $\Omega = \sqrt{V^2 + M_-^2}$.

One can see that all components of the evolution operator behave like
$e^{-i\lambda_{\pm}t}$, with $\lambda_{\pm} = M_+ \pm \Omega $. Since the
eigenfrequencies $\lambda_{\pm}$ are complex, the norm of the states is not
preserved in time.

A somewhat similar form of the $2 \times 2$ Hamiltonian matrix is responsible
for the oscillation of neutral kaons \cite{OKUN84}. The main difference between
the oscillation of neutral atoms and that of kaons is the mixing, which is
maximum for  kaons and exceedingly small for  atoms.

The Hamiltonian in Eq.~(\ref{HAMI}) also describes the effect of
neutron-antineutron oscillation in nuclear matter \cite{DGR85,MIK96}. In this
case, $M_i$ and $M_f$ would be the neutron and antineutron masses, $V$  the
baryon number violating potential, and $\Gamma $  the antineutron width related
to  annihilation channels (in  vacuum $M_i = M_f$ and $\Gamma = 0$).

The formalism  is also similar to the one used for describing oscillations and
decays of unstable neutrinos~\cite{Gonz08}.

\subsection{Oscillations of two stable atoms}

If the two atoms are stable, then $\Gamma = 0$ and $\Omega = (M_i - M_f)/2$.
The transition probability  is determined by the off-diagonal matrix element of
the evolution matrix (\ref{EVOLMATR}):
\begin{eqnarray}
|<f| e^{-i H_{\mathrm{eff}} t} |i>|^2 = \frac{V^2 }{\Omega^2} \sin^2 (\Omega
t).
\end{eqnarray}
This is just the case of oscillations of a two-level system described for
instance in Ref.~\cite{LLQM}. If $\Omega t \ll 1$, the transition probability
$\sim V^2 t^2$ is determined by the potential $V$ only. However, the exposure
time of atoms in double $\beta$-decay experiments (months and years) is 
greater than $1/\Omega$ by many orders of magnitude. By taking the average over
one period, we one arrives at
\begin{eqnarray}
|<f| e^{-i H_{\mathrm{eff}} t} |i>|^2 \approx \frac{2V^2}{(M_i-M_f)^2}.
\end{eqnarray}

In the transitions $(A,Z) \leftrightarrow (A,Z \pm 2)$, the composition of
valence electron shells changes and, thus, the chemical properties of the
substance. This circumstance can in principal  be used for registering the
oscillations of atoms. However, the potential $V$ is at least  30 orders of
magnitude smaller than the atomic mass difference.  For a hypothetical mass
difference of $M_i-M_f \approx 10$ keV one  finds \mbox{$|<f|e^{-i
H_{\mathrm{eff}} t}|i>|^2 < 10^{-60}$.}  Since degenerate ground-state masses
do not exist, this scenario is purely academic, and we turn to systems of a
stable mother and an excited daughter atom.

\subsection{Oscillations and resonant transitions between ground state and
excited atoms}

According to the arguments in Ref.~\cite{SIM08}, we assume a potential strength
of $V \sim 10^{-24}$ eV, a typical decay width of $ \Gamma \sim 1$ eV for a
medium-heavy atom, and a typical mass difference of $( M_i-M_f) \sim 1$ MeV. In
the lowest order in $ V $, we obtain
\begin{eqnarray}
\lambda_{+} &=& M_i + \Delta M - \frac{i}{2} \Gamma_1 ,  \label{LP} \\
\lambda_{-} &=& M_f - \frac{i}{2}\Gamma - \Delta M + \frac{i}{2} \Gamma_1,
\label{LM}
\end{eqnarray}
where
\begin{eqnarray}
\Delta M &=& \frac{V^2(M_i - M_f)}{(M_i - M_f)^2 + \frac{1}{4} \Gamma^2},
\label{M1} \\
\Gamma_1 &=& \frac{V^2 \Gamma }{(M_i - M_f)^2 + \frac{1}{4} \Gamma^2}.
\label{G1}
\end{eqnarray}
Since the width $\Gamma_1$ is small, the imaginary parts of 
$\lambda_{\pm}$ in Eqs.~(\ref{LP}) and (\ref{LM}) are negative, therefore, the states decay. Equation
(\ref{G1}) gives the decay rate of the initial atom in agreement with the
Breit-Wigner formula.

The excited atom manifests itself as a resonance in the decay amplitude. The
most favorable conditions for the detection of the violation of  lepton number
conservation occur in the transitions $(A,Z) \to (A,Z-2 )^{*}$, where the
masses of the initial and final states are equal (degenerate) and the decay
width of the daughter atomic nucleus is small.

The amplitude of finding the initial atom  $t$  seconds after its preparation
in the same initial state is determined by the diagonal matrix element of the
evolution operator:
\begin{eqnarray}
<i| e^{-i H_{\mathrm{eff}} t} |i> = e^{-i\lambda_- t}\frac{V^2}{4 M_-^2} +
e^{-i\lambda_+ t}(1 - \frac{V^2}{4 M_-^2}).  \label{II}
\end{eqnarray}
The second term oscillates with the frequency $\approx M_i$ and decays with the
rate $\Gamma_1$. At a low frequency, the system is unable to return to the
initial state and decays. In this case one can talk about a lepton number
violating decay of the initial atom with a width $\Gamma_1$ (Eq.~\ref{G1}).

The decay width $\Gamma_1$ reaches the unitary limit
\begin{eqnarray}
\Gamma_1^{\mathrm{max}}= \frac{4 V^2}{\Gamma} \label{unitarylimit}
\end{eqnarray}
in the case of a complete degeneracy between initial and final state. From  an
experimental standpoint, where one would search for such lepton number
violating decays, this would be the case of highest interest.

\section{Analysis of $0\nu$ECEC half-lives throughout periodic table}
\setcounter{equation}{0}

The selection of atomic systems with the potentially shortest $0\nu$ECEC
half-lives is based on  equation~(\ref{G1}).
The equation shows that the decay rate is determined by three quantities: the
mass difference between the initial and final states,
the decay width of the final state, and the transition potential.

The mass difference depends on the Q-value of the ECEC decay and the energy of
the two electron vacancies in the final atom.
In atomic physics, the electron binding energies can usually be calculated with
an accuracy of several eV. We borrowed the binding
energies from Ref.~\cite{LARK}. Noticeable corrections will, however, arise
from the Coulomb interaction of the two holes. These calculations
are carried out on the basis of the Dirac equation taking into account the
screening effects of the nuclear charge.

The decay width of the final atom is determined by the dipole emission rate
leading to the de-excitation of the electron shell.
{In the nonrelativistic approximation,} the capture
rates  of  two electrons from the higher shells, like L, M, or N shells,  scale
with the principal
quantum numbers $n_1,n_2$ as $(\alpha Z/n_1)^3(\alpha Z/n_2)^3$, but we will
see that in the unitarity limit
of a resonant decay this strong reduction of the probability could possibly
{be compensated by smallness of the de-excitation rates.}
The total decay width of the system is given by the sum of the widths of the
atomic  and the  nuclear state. In most cases the decay width
of the excited nucleus is  smaller than the one of the atomic state by at least
an order of magnitude and can, therefore, be neglected.
{The process of Auger electron emission as the alternative 
de-excitation process of the atom is also taken into account
following the results of Ref. \cite{CAMP}.
The Auger electron emission is faster than the electromagnetic decays for low
atomic $Z$.}

The transition potential contains the uncertainties of the transition matrix
elements connected with the complicated structure of the nuclear excitation. 
In order to obtain numerical estimates, we factorize the $0\nu$ECEC matrix element on a product of the atomic
physics factor and the nuclear matrix element. This simplification is justified due to weak radial dependence
of the $s_{1/2}$ and $p_{1/2}$ electron wave functions inside nuclei. Further, we normalize all the $0\nu$ECEC
nuclear matrix elements to the value of nuclear matrix element obtained for the ground state to ground state transition 
$^{152}_{\,\,64}$Gd$ \to ^{152}_{\,\,62}$Sm in Ref. \cite{erice11}. 
The contributions from the electron shell are determined by different combinations of the relativistic
wave functions of electrons for the capture from different shells.
We systematically examine transitions between all the states
$|J_f - J_i| = 0,1,2,\ldots$ with the parities $\pi_f \pi_i = \pm
1$
for the capture of two electrons with orbital angular momenta $0 \leq l_1 + l_2
\leq 2$
and the principal quantum numbers $1 \leq n_1,n_2 \leq 4$.

\subsection{Coulomb interaction energy of electron holes}

The binding energy of electrons in the inner atomic shells varies from $10$ eV
in light nuclei up to $100$ keV in heavy nuclei. In the
outer shells, the binding energy is a few eV, both in light and heavy nuclei. 
Since  electrons are usually captured from the
most inner shells, the electron binding energy gives sizeable contribution to
the energy balance in the double electron capture.
We use data of electron binding energies reported by Larkins~\cite{LARK}. Those
are accurate to better than an eV for
light nuclei and to a few tens of eV in heavy nuclei. The relevant binding
energies in the context of ECEC are, of course,
always those for the final daughter atom with $Z-2$.

In heavy elements, the electron hole interaction energies may reach values of a
few keV (for  a double K-shell vacancy), which
is quite large for our problem in question. It is therefore 
{essential} to calculate the interaction energies
of two electron holes and include them into the total energy equation.

We used an approach that takes screening of the Coulomb potential by electrons
occupying other orbitals into account.
The {shielding effect} can be estimated from the known energy
$\varepsilon$ of the bound electrons. In the non-relativistic theory, the
effective charge $Z_{*}$ can be found from equation
\begin{equation}
\varepsilon = m - \frac{\alpha^2Z_{*}^{2}}{2n^{2}}m, \label{atom}
\end{equation}
where $m$ is the electron mass, $n$ is the principal quantum number. In the
non-relativistic theory, the electron velocity $v \sim \alpha Z/n$  increases
with the nuclear charge, which requires a relativistic treatment for heavy
nuclei.

The binding energies in the Coulomb field are known from the Dirac relativistic
wave equation \cite{BERE}. Given $\varepsilon $, the effective charge may be
found from
\begin{equation}
\alpha ^{2}Z_{*}^{2} = \frac{\lambda^2}{m^2} \left( \kappa ^{2} + n_{r}^{2} -
2n_{r}\frac{n_{r}\lambda^2 - \varepsilon \sqrt{-n_{r}^{2}\lambda^2 +\kappa
^{2}m^2}}{m^2}\right),  \label{ZSRE}
\end{equation}
where $\lambda = \sqrt{m^2 - \varepsilon^2}  $, $\kappa = - (2j + 1)(j - l)$,
$j = l \pm 1/2$, and $n_{r} = n - (j + 1/2)\geq 0$. Near the limit of
$\varepsilon \to m$, the non-relativistic formula is recovered.

The effective charge $Z_{*}<Z$ takes into account the screening of the Coulomb
potential, as well as the finite nuclear size. Given that the electron-shell
wave functions are known, one can calculate the interaction energy of electron
holes.

We consider transitions between nuclei with good quantum numbers $J^{\pi}$, so
the two-electron wave function should have good total angular momentum $J$, 
projection $M$, and  parity. This can be arranged by weighting the two-electron
wave function with the Clebsch-Gordan coefficients
\begin{equation}
\psi _{\beta \delta }^{JM}(\mathbf{x}_{1},\mathbf{x}_{2})= \sum_{m_{\beta}
m_{\delta }}C_{j_{\beta }m_{\beta }j_{\delta }m_{\delta }}^{JM} \Psi _{\beta
m_{\beta }}(\mathbf{x}_{1})\Psi _{\delta m_{\delta }}(\mathbf{x}_{2}).
\label{CG}
\end{equation}
Here, $\Psi _{\alpha m_{\alpha}}(\mathbf{x})$ ($\alpha =(njl)$) is the
relativistic wave function of the electron in the Coulomb field.

The wave function of two electrons can be written as follows:
\begin{equation}
\Psi _{\beta \delta }^{JM}(\mathbf{x}_{1},\mathbf{x}_{2})=\frac{1}{\sqrt{2}}
(\psi _{\beta \delta }^{JM}(\mathbf{x}_{1},
\mathbf{x}_{2})-(-)^{j_{\beta}+j_{\delta }-J}\psi _{\delta \beta }^{JM}
(\mathbf{x}_{1},\mathbf{x}_{2})). \label{TWOE}
\end{equation}
The interaction energy of two electron holes can be obtained from equation
\begin{equation}
\epsilon_{C}=\int d\mathbf{x}_{1}d\mathbf{x}_{2}\Psi _{\beta \delta }^{JM\dagger}
(\mathbf{x}_{1},\mathbf{x}_{2})\frac{e^{2}}{|\mathbf{x}_{1}-\mathbf{x}_{2}|}
\Psi_{\beta \delta }^{JM}(\mathbf{x}_{1},\mathbf{x}_{2}). \label{ECRE}
\end{equation}
The case of two holes with identical quantum numbers $\alpha = \beta$ and
$m_{\alpha}\ne m_{\beta}$ requires special attention. The states $J=2j$
$\mathrm{mod}(2)$ are symmetric over the $m_{\alpha} \leftrightarrow 
m_{\beta}$ permutations and do not exist. In the $J=2j+1$ $\mathrm{mod}(2)$
case the states are antisymmetric over the $m_{\alpha} \leftrightarrow 
m_{\beta}$ permutations, and the interaction energy should be divided by a
factor of $2$, since the superposition (\ref{CG}) changes the overall
normalization of the two-electron wave function, as it follows from
$C_{jm_{1}jm_{2}}^{JM}=(-1)^{J-2j}C_{jm_{2}jm_{1}}^{JM}$.

To simplify notations, we label the final states by indices $\alpha ,\gamma$
and the initial ones by $\beta ,\delta$. The labels take values $1,2$ to
indicate first and second electrons. There are two possibilities for the final
state, $(\alpha,\gamma) = (1,2)$ and $(2,1)$, and two possibilities for the
initial state, $(\beta,\delta) = (1,2)$ and $(2,1)$.

Equation (\ref{ECRE}) can be written as
\begin{equation}
\epsilon_{C} = K_{\beta \delta \beta \delta }^{JM}-(-)^{j_{\beta
}+j_{\delta}-J} K_{\beta \delta \delta \beta }^{JM},
\end{equation}
where
\begin{equation}
K_{\alpha \gamma \beta \delta }^{JM}=\sum_{m_{\alpha }m_{\gamma
}m_{\beta}m_{\delta }} C_{j_{\alpha }m_{\alpha }j_{\gamma }m_{\gamma}}^{JM}
C_{j_{\beta }m_{\beta }j_{\delta }m_{\delta }}^{JM}K_{\alpha m_{\alpha }\gamma
m_{\gamma } \beta m_{\beta } \delta m_{\delta }}
\end{equation}
and
\begin{eqnarray*}
~K_{\alpha m_{\alpha }\gamma m_{\gamma } \beta m_{\beta } \delta m_{\delta }}
=\int d\mathbf{x}_{1}d\mathbf{x}_{2}\left[ \Psi _{\alpha m_{\alpha
}}^{\dagger}(\mathbf{x}_{1})\Psi _{\beta m_{\beta }}(\mathbf{x}_{1})\right]
\frac{e^{2}}{|\mathbf{x}_{1}-\mathbf{x}_{2}|}\left[ \Psi _{\gamma m_{\gamma
}}^{\dagger} (\mathbf{x}_{2})\Psi _{\delta m_{\delta }}(\mathbf{x}_{2})\right].
\label{EC}
\end{eqnarray*}
Further simplifications appear after the use of the expansion
\begin{equation}
\frac{1}{|\mathbf{x}_{1}-\mathbf{x}_{2}|}=\sum_{lm}\frac{4\pi }{2l+1}
\frac{r_{<}^{l}}{r_{>}^{l+1}}Y_{lm}(\mathbf{n}_{1})Y_{lm}^{*}(\mathbf{n}_{2}),
\end{equation}
where $r_{<}=\min (r_{1},r_{2})$, $r_{>}=\max (r_{1},r_{2})$, and
$\mathbf{n}_{i}$ are unit vectors toward $\mathbf{x}_{i}$.

\begin{figure} [t] 
\begin{center}
\includegraphics[width = 0.62\textwidth]{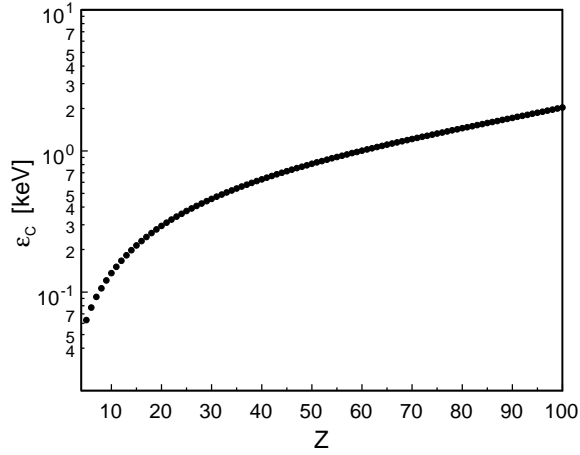}
\caption{Coulomb interaction energy of two $1s$ electron holes versus nuclear
charge $Z$.} 
\label{fig:2}
\end{center}
\end{figure}

The angular integrals 
are calculated with the use of equation
\begin{equation}
\int d\Omega _{\mathbf{n}}\Omega _{\alpha m_{\alpha}}^{\dagger}(\mathbf{n})\Omega
_{\beta m_{\beta}} (\mathbf{n})Y_{lm}(\mathbf{n})= C_{j_{\beta }m_{\beta
}lm}^{j_{\alpha}m_{\alpha }}\mathcal{A}_{\alpha \beta }^{l},
\end{equation}
where $\Omega _{\alpha m_{\alpha} }(\mathbf{n})$ are spherical spinors \cite{LLQM},
\begin{equation}
\mathcal{A}_{\alpha \beta }^{l}=(-)^{1/2+j_{\beta }+l_{\alpha }+l}
\sqrt{\frac{[l][l_{\beta }][j_{\beta }]}{4\pi }} C_{l_{\beta
}0l0}^{l_{\alpha}0}\left\{
\begin{array}{lll}
1/2 & l_{\beta } & j_{\beta } \\
l & j_{\alpha } & l_{\alpha }
\end{array}
\right\}
\end{equation}
and $[x]=2x+1$.

The remaining two-dimensional integral over the radial variables,
\begin{equation}
K_{\alpha \gamma \beta \delta }^{JM}=e^{2}\sum_{l} \frac{4\pi }{2l+1}
\mathcal{C}_{\alpha \beta \gamma \delta }^{Jl}\int
r_{1}^{2}dr_{1}r_{2}^{2}dr_{2}\frac{r_{<}^{l}}{r_{>}^{l+1}} \mathcal{F}_{\alpha
\beta }^{l}(r_{1})\mathcal{F}_{\gamma \delta }^{l}(r_{2}),
\end{equation}
can be calculated numerically or analytically using Maple. Here,
\begin{equation}
\mathcal{C}_{\alpha \beta \gamma \delta }^{Jl}=(-)^{j_{\gamma }+j_{\beta }+J}
\sqrt{[j_{\alpha }][j_{\gamma }]}\left\{
\begin{array}{lll}
l & j_{\gamma } & j_{\delta } \\
J & j_{\beta } & j_{\alpha }
\end{array}
\right\}
\end{equation}
and
\begin{equation}
\mathcal{F}_{\alpha \beta }^{l}(r)=f_{\alpha }(r)f_{\beta }(r)
\mathcal{A}_{\alpha \beta }^{l}+g_{\alpha ^{\prime }}(r) g_{\beta ^{\prime
}}(r)\mathcal{A}_{\alpha ^{\prime }\beta ^{\prime }}^{l},
\end{equation}
where $\alpha = (njl)$, $\alpha^{\prime} = (njl^{\prime})$, and $l^{\prime} =
2j - l$, and similarly for $\beta$.

In Fig. \ref{fig:2} we show the Coulomb interaction energy of two K holes as a
function of the nuclear charge.
As expected, the interaction energy increases approximately linearly with $Z$.

More accurate estimates of the interaction energy of electron holes can be
obtained on the basis of the Breit potential,
which takes into account relativistic effects $\sim (v/c)^2$. The accuracy of
the present estimate, therefore, can be evaluated
to be of order  $(\alpha Z/n)^2$. For $Z = 60$ the expected error in the
interaction energy of two K electron holes is about 20\%.
The accuracy is, of course, better for higher-shell atomic excitations.

\subsection{Natural widths of excited electron shells}
Decays of  excited atoms are dominated by electric dipole transitions with the
emission of X-ray photons and/or the emission of
Auger electrons. Dipole decays are described in the literature in detail (see,
e.g., \cite{BESA}). For K vacancies, characteristic X-ray
dipole emissions are dominant in atoms with $Z \gtrsim 35$. The  dipole
transition $2p \to 1s$ has a probability
$\Gamma = 4 \times 10^{-7} Z^4$ eV (see, e.g., \cite{BESA}). Since for a
$(n+1)p \to ns$ decay, $\omega_{\alpha\beta } \sim 1/n^3$
and $ d_{ \alpha\beta } \sim n^2$, the corresponding width decreases as
$\Gamma_{\alpha\beta } \sim 1/n^5$ for transitions from higher orbits.
This effect is of interest, since the unitary limit of the width of the lepton
number violating decay is inversely proportional to the width
of the daughter atom (see Eq.~(\ref{unitarylimit})).

Auger transitions  of excited atoms with one electron vacancy  are well studied
theoretically and experimentally (for a review see \cite{CAMP}).
{
The width of a two-hole state $\alpha\beta$ is represented by
\begin{equation}
\Gamma_{\alpha\beta} =  \Gamma_{\alpha}+ \Gamma_{\beta} + \Gamma^{*},
\label{GAMMAT}
\end{equation}
where $\Gamma^{*}$ is the de-excitation width of daughter nucleus.}
Numerical values of the one-hole widths
$\Gamma_{\alpha}$ are taken from Ref. \cite{CAMP}. These values cover the range
$10 \leq Z \leq 92$ and one-electron vacancies from K to N7 shells
($1 \leq n \leq 4$).
Equation (\ref{GAMMAT}) neglects the contributions from two-hole correlations.

\subsection{Lepton number violating potential}

We consider the $0\nu$ECEC process assuming the standard form of the
$\beta$-decay Hamiltonian
\begin{equation}
{\cal H}^\beta(x)=\frac{G_{\beta}}{\sqrt{2}} \bar{e} (x)\gamma^{\mu} (1 -
\gamma_5) \nu_{e}(x) J_\mu (x) + \mathrm{h. c.},
\end{equation}
where $G_\beta = G_F \cos{\theta_C}$ and  ${\theta_C}$ is the Cabibbo angle.
The field operators of the  electron and  electron neutrino are denoted as
$e(x)$ and $\nu_{e}(x)$.

The left-handed electron neutrino $\nu_{e L}(x) = \frac{1-\gamma_5}{2}
\nu_e(x)$ is a superposition of the left-handed projections of Majorana
neutrinos $\chi_k$ with diagonal masses $m_k$:
\begin{equation}
\nu_{e L}(x) = \sum^3_{i=1} U_{e k} \chi_{k L}(x).
\end{equation}
Here, $U$ is PMNS neutrino mixing matrix. Majorana neutrinos are truly neutral
particles and obey $C {\overline{\chi}}_k^T = \xi_k \chi_k$, where $C =
i\gamma_2 \gamma_0$ is the charge conjugation matrix and $\xi_k$ is the phase
factor.

The strangeness conserving charged hadron current has the form
\begin{equation}
J^{\mu} (x)=\bar{p}(x)\gamma^{\mu} (g_{{V}} - g_{{A}} \gamma_{{5}}) n(x),
\end{equation}
where $n(x)$ and $p(x)$ are the field operators of the neutron and the proton,
and the vector and axial-vector coupling constants are $g_{{V}}=1$ and
$g_{{A}}=1.25$.

The potential of the $0\nu$ECEC capture of two electrons with the total angular
momentum $J$ and projection $M$ can be written as follows
\begin{eqnarray}
V_{\alpha\beta}(J^{\pi}_f) &=&  i m_{\beta\beta} \left(\frac{G_\beta}{\sqrt{2}}\right)^2
\frac{1}{\sqrt{1 + \delta_{\alpha\beta}}}
                              \sum_{m_\alpha m_\beta} C^{JM}_{j_\alpha m_\alpha
j_\beta m_\beta} \int d{\mathbf{x}}_1 d{\mathbf{x}}_2 \label{LNVP1}\\
&\times& {\Psi_{\alpha m_\alpha}}^T ({\mathbf{x}}_1) C \gamma^\mu \gamma^\nu (1 -
\gamma_5 )  {\Psi_{\beta m_\beta}} ({\mathbf{x}}_2)
\int \frac{e^{-i \vec{q}\cdot({\mathbf{x}}_1-{\mathbf{x}}_2)}}{2 q_0}  \frac{d
\vec{q}}{(2\pi)^3} \nonumber \\
&\times& \sum_n \left[ \frac{<A,
Z-2|J_\mu({\mathbf{x}}_1)|n><n|J_\nu({\mathbf{x}}_2)|A,Z>} {q_0+  E_n - M_i -
\varepsilon_\beta } \right. \nonumber \\  
&+& \left. \frac{<A, Z - 2|J_\nu({\mathbf{x}}_2)|n><n|J_\mu({\mathbf{x}}_1)|A,Z>} {q_0+ E_n  - M_i -
\varepsilon_\alpha } \right] - (\alpha \leftrightarrow \beta ).
\nonumber
\end{eqnarray}
Here, $J_\mu (\mathbf{x})$ is the weak charged current in the Heisenberg
representation. $|A,Z>$ and $|A,Z - 2>$ are states of the initial and final
nuclei. The sum is taken over all excitations of the intermediate nucleus $(A,Z
- 1)$. $\Psi_{\alpha m_\alpha}({\mathbf{x}})$ is a wave function of the bound
electron with quantum numbers $\alpha =(n_\alpha j_\alpha l_\alpha)$,
projection of the total angular momentum $m_\alpha$, and energy
$\varepsilon_\alpha$. The factor ${1}/{\sqrt{1 + \delta_{\alpha\beta}}}$ takes
statistics of the captured electrons into account: $\delta_{\alpha \beta} = 1$
for the identical states and $\delta_{\alpha \beta} = 0$ otherwise.

In the derivation of $V_{\alpha\beta} (J^\pi)$, we neglected the small neutrino
masses ($m_i \ll 10$ eV) in the neutrino potential, since the average exchange
momentum in the process is large, $|\vec{q}| \simeq 200$ MeV/c. Further
simplifications are as follows:
\begin{itemize}
\item[i)] Non-relativistic impulse approximation for the nuclear current:
\begin{equation}\label{NRIA}
J^{\mu} (0,\mathbf{x}) = \sum_{n=1}^{A} \tau^-_n [g_V g^{\mu 0} + g_A 
(\sigma_k)_n g^{\mu k} ] \delta(\mathbf{x}-{\mathbf{x}}_n).
\end{equation}
\item[ii)]  Closure approximation for the intermediate states: The excitation
energies of the intermediate states $E_n - M_i$ are replaced by an average
value $<E> \approx 8$ MeV. In addition, we set $\varepsilon_{\alpha , \beta}
\approx m$.
The sum entering Eq.~(\ref{LNVP1}) is then calculated using completeness
condition $\sum_n |n><n| = 1$.\\
\item[iii)] We restrict the calculation of the Majorana neutrino exchange
potentials to the most favorable cases of even-even nuclei. Then the angular
momentum of the initial nucleus is $0^+$ and the angular momentum of the final
(possibly excited) nucleus $J^\pi$ must be balanced by the capture of the
atomic electrons and the angular momentum of the atomic state.
\end{itemize}

\noindent The potential can finally be written as
\begin{eqnarray}
V_{\alpha \beta} (J_f^{\pi}) = \frac{1}{4 \pi}~ G^2_{\beta} m_{\beta\beta}
\frac{g^2_A}{R} \sqrt{2 J_f + 1} {\cal M}_{\alpha\beta}(J_f^{\pi}).
\label{LNVP2}
\end{eqnarray}
In the case of a capture of $s_{1/2}$ and $p_{1/2}$ electrons and of a
favorable case for the nuclear transitions $0^+ \to J_f^{\pi} =
0^{\pm},1^{\pm}$, the matrix elements ${\cal M}_{\alpha\beta}(J_f^{\pi})$ are
given in Appendix B.

The numerical analysis of the $0\nu$ECEC transition is performed by factorizing
the electron shell structure and the nuclear matrix element:
\begin{eqnarray}
{\cal M}_{\alpha\beta}(J_f^{\pi}) \approx {\cal A}_{\alpha\beta} M^{0\nu}(J_f^{\pi}). \label{LNVP3}
\end{eqnarray}
Here, ${\cal A}_{\alpha\beta}$ are lepton factors averaged over the nuclear volume. For low-$J$ transitions, 
the lepton factors are given in Table 1.

\begin{table}[tbp]
\centering 
\renewcommand{\arraystretch}{1.1}
\caption{Combinations of the averaged upper and lower bi-spinor
components of the electron wave functions entering the $0\nu$ECEC potential for
transitions $0^{+}_i \rightarrow 0^{\pm}_f,1^{\pm}_f$. Here, $\alpha,\beta$ are
quantum numbers of electron hole states, and $F^{(\pm) }_{\alpha
\beta}(r_n,r_m)$ and $H^{(\pm) }_{\alpha \beta}(r_n,r_m)$ are defined in
Appendix B.
If electron holes are in the same state,
${\cal A}_{\alpha\beta}$ should be divided by an additional factor $\sqrt{2}$.}
\label{tab:table1}
\vspace{2mm}
\begin{tabular}{|c|c|}
\hline \hline
Transitions            & $ {\cal A}_{\alpha\beta}$  \\
\hline \hline
$0^+ \to 0^+$    & $ < F^{(+) }_{\alpha \beta}(r_n,r_m)>$ \\
$0^+ \to 0^-$    & $ < H^{(+) }_{\alpha \beta}(r_n,r_m)>$ \\
$0^+ \to 1^+$    & $ < F^{(-)2}_{\alpha \beta}(r_n,r_m) >^{1/2}$ \\
$0^+ \to 1^-$    & $ < (H^{(-)}_{\alpha \beta}(r_n,r_m) - H^{(-)}_{\alpha
\beta}(r_m,r_n))^2/4>^{1/2}$ \\
\hline \hline
\end{tabular}
\end{table}

\noindent The nuclear matrix elements of $0^+ \to 0^{\pm}$ transitions have the
form
\begin{eqnarray}
M^{0\nu}(0^+_f) &=& <0^+_f \parallel \sum_{nm} \tau^-_n \tau^-_m  h(r_{nm})
[-\frac{g^2_V}{ g^2_A} +({\mbox{\boldmath{$\sigma$}}}_n \cdot {\mbox{\boldmath{$\sigma$}}}_m) ] \parallel
0^+_i>, \label{ZEPL}\\
M^{0\nu}(0^-_f) &=& <0^-_f \parallel \sum_{nm} \tau^-_n \tau^-_m  h(r_{nm})
({\hat{\mathbf{r}}}_n-{\hat{\mathbf{r}}}_m) \nonumber \\
&\times& [\frac{g_V}{g_A}  (\mbox{\boldmath{$\sigma$}}_n-\mbox{\boldmath{$\sigma$}}_m) -
i (\mbox{\boldmath{$\sigma$}}_n\times\mbox{\boldmath{$\sigma$}}_m)]
\parallel 0^+_i>, \label{ZEMI}
\end{eqnarray}
where
\begin{equation}
h(r_{nm}) = \frac{2}{\pi} R \int_0^\infty j_0(q r_{nm}) \frac{q_0}{q_0 + <E> -
m} dq. \label{npot}
\end{equation}
We note that the nuclear matrix element $M^{0\nu}(0^+_f)$ also
appears  in the calculation of the $0\nu\beta\beta$-decay process
\cite{dbdreviews}.

For the transition $0^+ \to 1^{\pm}$ the lepton parts and the nuclear matrix
elements are evaluated as discussed in Appendix~A.
We note that ${\cal A}_{\alpha\beta}$ vanishes for $0^+ \to 1^{\pm}$
transitions whenever the two electrons are captured from states with
the same quantum numbers $(njl)$ {for $j=1/2$}. This is the reason,
why e.g. the transition $^{162}_{ 68}$Er$(0^+) \to ^{162}_{ 66}$Dy$^{**}$
(1$^+$, $E_{\rm nucl}^x=1745.72$ keV) is excluded from the analysis. If 
electrons are captured from different states, e.g., two $s_{1/2}$ electrons
from different shells ($n_\alpha \ne n_\beta$) or from $s_{1/2}$ and $p_{1/2}$
states, the transition $0^+ \to 1^{\pm}$ is allowed and is considered in our
analysis.

The dominant combinations of upper and lower component of bi-spinors, which
enter the lepton part of the matrix elements after the factorization, are
listed in Table \ref{tab:table1}. The definition of functions $F^{(\pm)
}_{\alpha \beta}(r_n,r_m)$ and $H^{(\pm) }_{\alpha \beta}(r_n,r_m)$ can be
found in Appendix B. The decay rates of other transitions are estimated roughly
with
\begin{equation}
2\pi \sqrt{2} {\cal A}_{\alpha\beta} \sim \sqrt{ <f_\alpha^2 + g_\alpha^2>
       <f_\beta ^2 + g_\beta ^2>}.
\label{wfappox}
\end{equation}

The parity non-conservation in the weak interactions allows for instance
transitions $0^+_i \rightarrow 0^-_f,1^-_f$ accompanied by
{capture of two $S$-wave or two $P$-wave electrons}.

\subsection{Likely resonant $0\nu$ECEC transitions}

We have considered all the nuclei and their excited states registered in the
database of the Brookhaven National Laboratory \cite{BNL} in August 2010,
as well as all the combinatorial possibilities associated with the capture of
two electrons. The selection criteria are as follows:
\begin{itemize}
\item[i)] The excitation energies are usually known with precision much higher
than the atomic ground-state masses. We selected those pairs, where degeneracy
occurs within the bounds given by a three standard deviation error  of the
ground-state mass measurements.

\item[ii)] The unitary limit for the normalized half-life is less than
$10^{27}$ years. 
\end{itemize}

\noindent 
The half-lives are calculated using the formula
\begin{equation}
T_{1/2} = \frac{\ln 2}{\Gamma_1}, \label{T}
\end{equation}
where $\Gamma_1$ is given by Eq.~(\ref{G1}).

Tables~\ref{tab:table2},~\ref{tab:table3},~and~\ref{tab:table44} present the
results of the selection { over stable parent isotopes}.
We show the  natural abundances NA, the spin-parity of the final nucleus
$J^{\pi}_f$, the excitation energy of the final nucleus
$E_{\rm nucl}^x = M^*_{A,Z-2} - M_{A,Z - 2}$, and the total mass difference
$M^{**}_{A,Z-2} - M_{A,Z}$. The two errors indicate the errors
of the ground-state mass measurements.  Shown  are as well the quantum numbers
of the two hole states $\alpha$ and $\beta$ in the electron shell,
the energy of the holes (not including the electron rest mass), the Coulomb
interaction energies of the holes $\epsilon_C$,
and the decay widths $\Gamma_{\alpha \beta}$.  The last two columns show the
minimum and maximum normalized $0\nu$ECEC half-lives.

Tables~\ref{tab:table2},~\ref{tab:table3},~and~\ref{tab:table44}
list all, but no more than 5 transitions with the lowest quantum numbers of
electron holes
for each pair of the elements.
If the spin is not fully determined, we took its lowest suggested value.

Rigorous calculations of the nuclear matrix elements (NME) based on the
structure of nuclear states have not yet been performed.
The objective here is to  first select promising pairs of nuclei on the basis
of rough estimates of the matrix elements, as these won't significantly
change the global picture. 
The half-lives are normalized to the nuclear matrix element of $\mathcal{M}^{0\nu}(0_f^{+}) = 6$, 
which roughly corresponds to the maximum evaluated value of NMEs for medium-heavy nuclei \cite{src}. 
Transitions to excited states are suppressed due to dissimilarity of the nuclear wave functions \cite{suhonen,Kolh11}. 

\begin{landscape}

\begin{table}[]
\vspace{-4mm}
\caption{Likely resonant $0\nu$ECEC transitions. The list contains only
those initial and final atoms, which are potentially degenerate at the level
of a 3$\sigma$ experimental error, and for which the unitary limit of the
half-lives ${\tilde T}_{1/2}^{\min}$ is below $10^{27}$ years.
The first column shows  the natural abundances (NA in \%) of the parent nuclei.
In column 3 the spin and parity of the excited final nuclei (or their suggested values 
in parentheses or, if unknown, their assumed values in square brackets) are listed. 
Column 4 shows the excitation energies of the final nuclei with their present experimental errors.  
Column 5 lists the total mass differences, including the hole energies of the final nuclei, 
where the first error indicates the experimental uncertainty of the parent ground-state mass 
and the second the one of the daughter. The quantum numbers
of the electron holes in the next two columns are $(n2jl)$,
which are the principal quantum number, twice the total angular momentum, and
the orbital angular momentum. Columns 8, 9, and 10 list the hole
energies and the Coulomb interaction energies. Column 11 shows {the
widths} of the  excited electron shells. The last two columns show the minimum
and maximum half-lives of the $0\nu$ECEC transitions (in years). Masses,
energies and widths are given in keV.}
\label{tab:table2} 
\centering
\renewcommand{\arraystretch}{1.1}
\addtolength{\tabcolsep}{1pt}
\scriptsize
\vspace{2mm}
\begin{tabular}{|c|c|c|c|r|c|c|r|r|r|r|c|c|c|c|c|c|}
\hline \hline
NA  \T \B &   Transition  & $J^{\pi}_f$ & $M^*_{A,Z-2} - M_{A,Z - 2}$ &
$M^{**}_{A,Z-2} - M_{A,Z}$ & $(n 2j l)_{\alpha}$ & $(n 2j l)_{\beta}$ &
$\epsilon_{\alpha}^*\;\;$ & $\epsilon_{\beta}^*\;\;$ & $\epsilon_{C}$ &
$\Gamma_{\alpha \beta}\;\;\;\;\;\;$  & $\tilde{T}_{1/2}^{\min}$ &
$\tilde{T}_{1/2}^{\max}$ \\
\hline \hline
5.52\% \T                                 &$^{ 96}_{ 44}$Ru$ \to ^{ 96}_{
42}$Mo$^{**}$  & 0$^{+ }$    &2742 $\pm$ 1      & 24.1$\pm$ 7.9$\pm$ 1.9    &  310
   &  410    &  0.50    &  0.06    &  0.02    &$9.5\times 10^{  -3}$   
&$3\times 10^{ 26}$       &$9\times 10^{ 33}$      \\
     \T \B                                   &                                      
    &             &                  & 23.7$\pm$ 7.9$\pm$ 1.9    &  410    & 
410    &  0.06    &  0.06    &  0.01    &$6.4\times 10^{  -3}$    &$1\times
10^{ 27}$       &$6\times 10^{ 34}$      \\
\hline
1.25\%   \T                               &$^{106}_{ 48}$Cd$ \to ^{106}_{
46}$Pd$^{**}$  &   $[0^+]$   &2737 $\pm$ 1      & 16.5$\pm$ 5.9$\pm$ 4.1    &  110
   &  110    & 24.35    & 24.35    &  0.74    &$1.3\times 10^{  -2}$   
&$3\times 10^{ 23}$       &$2\times 10^{ 30}$      \\
         \T                               &                                      
    &             &                  & -4.8$\pm$ 5.9$\pm$ 4.1    &  110    & 
210    & 24.35    &  3.60    &  0.23    &$1.0\times 10^{  -2}$    &$9\times
10^{ 23}$       &$3\times 10^{ 30}$      \\
         \T                               &                                      
    &             &                  & -5.1$\pm$ 5.9$\pm$ 4.1    &  110    & 
211    & 24.35    &  3.33    &  0.21    &$8.5\times 10^{  -3}$    &$1\times
10^{ 26}$       &$5\times 10^{ 32}$      \\
         \T                               &                                      
    &             &                  & -7.9$\pm$ 5.9$\pm$ 4.1    &  110    & 
310    & 24.35    &  0.67    &  0.07    &$1.4\times 10^{  -2}$    &$4\times
10^{ 24}$       &$9\times 10^{ 30}$      \\
         \T \B                            &                                      
    &             &                  & -8.5$\pm$ 5.9$\pm$ 4.1    &  110    & 
410    & 24.35    &  0.09    &  0.02    &$1.1\times 10^{  -2}$    &$7\times
10^{ 24}$       &$3\times 10^{ 31}$      \\
\hline
0.095\%  \T                               &$^{124}_{ 54}$Xe$ \to ^{124}_{
52}$Te$^{**}$  & [$0^+$]   &2853.2  $\pm$0.6  & -1.2$\pm$ 1.8$\pm$ 1.5    &  210  
 &  210    &  4.94    &  4.94    &  0.16    &$4.4\times 10^{  -3}$    &$2\times
10^{ 24}$       &$3\times 10^{ 30}$      \\
                                        &                                      
 \T   &             &                  & -1.6$\pm$ 1.8$\pm$ 1.5    &  210    & 
211    &  4.94    &  4.61    &  0.16    &$5.0\times 10^{  -3}$    &$2\times
10^{ 26}$       &$2\times 10^{ 32}$      \\
 \T                                       &                                      
    &             &                  & -5.2$\pm$ 1.8$\pm$ 1.5    &  210    & 
310    &  4.94    &  1.01    &  0.08    &$1.2\times 10^{  -2}$    &$9\times
10^{ 24}$       &$8\times 10^{ 30}$      \\
  \T                                    &                                      
       &             &                  & -5.4$\pm$ 1.8$\pm$ 1.5    &  210    & 
311    &  4.94    &  0.87    &  0.06    &$5.4\times 10^{  -3}$    &$5\times
10^{ 26}$       &$2\times 10^{ 33}$      \\
 \T  \B                                 &                                      
       &             &                  & -6.1$\pm$ 1.8$\pm$ 1.5    &  210    & 
410    &  4.94    &  0.17    &  0.02    &$4.6\times 10^{  -3}$    &$8\times
10^{ 24}$       &$7\times 10^{ 31}$      \\
\hline
0.185\%  \T  \B                             &$^{136}_{ 58}$Ce$ \to ^{136}_{56}$Ba$^{**}$  & 0$^{+ }$    &2315.32 $\pm$0.07 &-27.5$\pm$13.3$\pm$ 0.4    &  110
   &  110    & 37.44    & 37.44    &  0.93    &$2.6\times 10^{  -2}$   
&$1\times 10^{ 23}$       &$7\times 10^{ 29}$      \\
\hline
0.185\%  \T                             &$^{136}_{ 58}$Ce$ \to ^{136}_{56}$Ba$^{**}$  &  [0$^{+}$]  &2349.5  $\pm$0.5  &  6.7$\pm$13.3$\pm$ 0.4    &  110
   &  110    & 37.44    & 37.44    &  0.93    &$2.6\times 10^{  -2}$   
&$1\times 10^{ 23}$       &$2\times 10^{ 29}$      \\
         \T                                 &                                      
    &             &                  &-25.4$\pm$13.3$\pm$ 0.4    &  110    & 
210    & 37.44    &  5.99    &  0.30    &$1.5\times 10^{  -2}$    &$3\times
10^{ 23}$       &$4\times 10^{ 30}$      \\
        \T                                &                                      
    &             &                  &-25.8$\pm$13.3$\pm$ 0.4    &  110    & 
211    & 37.44    &  5.62    &  0.28    &$1.6\times 10^{  -2}$    &$3\times
10^{ 25}$       &$4\times 10^{ 32}$      \\
         \T                               &                                      
    &             &                  &-30.3$\pm$13.3$\pm$ 0.4    &  110    & 
310    & 37.44    &  1.29    &  0.09    &$2.4\times 10^{  -2}$    &$1\times
10^{ 24}$       &$1\times 10^{ 31}$      \\
         \T  \B                           &                                      
    &             &                  &-30.5$\pm$13.3$\pm$ 0.4    &  110    & 
311    & 37.44    &  1.14    &  0.08    &$1.7\times 10^{  -2}$    &$1\times
10^{ 26}$       &$2\times 10^{ 33}$      \\
\hline
0.185\%  \T                               &$^{136}_{ 58}$Ce$ \to ^{136}_{
56}$Ba$^{**}$  & $(1^+,2^+)$ &2392.1 $\pm$0.6   & 17.1$\pm$13.3$\pm$ 0.4    &  110
   &  210    & 37.44    &  5.99    &  0.21    &$1.5\times 10^{  -2}$   
&$8\times 10^{ 22}$       &$6\times 10^{ 29}$      \\ \T 
                                        &                                      
    &             &                  & 16.8$\pm$13.3$\pm$ 0.4    &  110    & 
211    & 37.44    &  5.62    &  0.29    &$1.6\times 10^{  -2}$    &$2\times
10^{ 24}$       &$2\times 10^{ 31}$      \\ \T
                                        &                                      
    &             &                  & 12.3$\pm$13.3$\pm$ 0.4    &  110    & 
310    & 37.44    &  1.29    &  0.07    &$2.4\times 10^{  -2}$    &$4\times
10^{ 23}$       &$9\times 10^{ 29}$      \\ \T
                                        &                                      
    &             &                  & 12.1$\pm$13.3$\pm$ 0.4    &  110    & 
311    & 37.44    &  1.14    &  0.08    &$1.7\times 10^{  -2}$    &$7\times
10^{ 24}$       &$3\times 10^{ 31}$      \\ \T \B
                                        &                                      
    &             &                  & 11.2$\pm$13.3$\pm$ 0.4    &  110    & 
410    & 37.44    &  0.25    &  0.03    &$1.6\times 10^{  -2}$    &$7\times
10^{ 23}$       &$3\times 10^{ 30}$      \\
\hline \hline
\end{tabular}
\end{table}

\end{landscape}

\begin{landscape}

\begin{table}[]
\vspace{-4mm}
\caption{Continued from Table \ref{tab:table2}.}
\label{tab:table3} \centering
\renewcommand{\arraystretch}{1.1}
\addtolength{\tabcolsep}{1pt}
\scriptsize
\vspace{2mm}
\begin{tabular}{|c|c|c|c|r|c|c|r|r|r|r|c|c|c|c|c|c|}
\hline \hline
     NA \T \B  &   Transition  & $J^{\pi}_f$ & $M^*_{A,Z-2} - M_{A,Z - 2}$ &
$M^{**}_{A,Z-2} - M_{A,Z}$ & $(n 2j l)_{\alpha}$ & $(n 2j l)_{\beta}$ &
$\epsilon_{\alpha}^*\;\;$ & $\epsilon_{\beta}^*\;\;$ & $\epsilon_{C}$ &
$\Gamma_{\alpha \beta}\;\;\;\;\;\;$  & $\tilde{T}_{1/2}^{\min}$ &
$\tilde{T}_{1/2}^{\max}$ \\
\hline \hline
0.185\%                                 &$^{136}_{ 58}$Ce$ \to ^{136}_{
56}$Ba$^{**}$  & $(1^+,2^+)$ &2399.87 $\pm$0.05 & 24.9$\pm$13.3$\pm$ 0.4    &  110
   &  210    & 37.44    &  5.99    &  0.21    &$1.5\times 10^{  -2}$   
&$8\times 10^{ 22}$       &$1\times 10^{ 30}$      \\ \T
                                        &                                      
    &             &                  & 24.6$\pm$13.3$\pm$ 0.4    &  110    & 
211    & 37.44    &  5.62    &  0.29    &$1.6\times 10^{  -2}$    &$2\times
10^{ 24}$       &$3\times 10^{ 31}$      \\ \T
                                        &                                      
    &             &                  & 20.0$\pm$13.3$\pm$ 0.4    &  110    & 
310    & 37.44    &  1.29    &  0.07    &$2.4\times 10^{  -2}$    &$4\times
10^{ 23}$       &$2\times 10^{ 30}$      \\ \T
                                        &                                      
    &             &                  & 19.9$\pm$13.3$\pm$ 0.4    &  110    & 
311    & 37.44    &  1.14    &  0.08    &$1.7\times 10^{  -2}$    &$7\times
10^{ 24}$       &$5\times 10^{ 31}$      \\ \T \B
                                        &                                      
    &             &                  & 18.9$\pm$13.3$\pm$ 0.4    &  110    & 
410    & 37.44    &  0.25    &  0.03    &$1.6\times 10^{  -2}$    &$7\times
10^{ 23}$       &$5\times 10^{ 30}$      \\
\hline \T
0.20\%                                  &$^{152}_{ 64}$Gd$ \to ^{152}_{ 62}$Sm$
^{*} $  & 0$^{+ }$    &   0              & -0.8$\pm$ 2.5$\pm$ 2.5    &  110    & 
210    & 46.83    &  7.74    &  0.34    &$2.3\times 10^{  -2}$    &$2\times
10^{ 23}$       &$8\times 10^{ 26}$      \\ \T
                                        &                                      
    &             &                  & -1.3$\pm$ 2.5$\pm$ 2.5    &  110    & 
211    & 46.83    &  7.31    &  0.32    &$2.3\times 10^{  -2}$    &$2\times
10^{ 25}$       &$2\times 10^{ 29}$      \\ \T
                                        &                                      
    &             &                  & -7.1$\pm$ 2.5$\pm$ 2.5    &  110    & 
310    & 46.83    &  1.72    &  0.11    &$3.2\times 10^{  -2}$    &$7\times
10^{ 23}$       &$1\times 10^{ 29}$      \\ \T
                                        &                                      
    &             &                  & -7.3$\pm$ 2.5$\pm$ 2.5    &  110    & 
311    & 46.83    &  1.54    &  0.10    &$2.5\times 10^{  -2}$    &$5\times
10^{ 25}$       &$2\times 10^{ 31}$      \\ \T \B
                                        &                                      
    &             &                  & -8.5$\pm$ 2.5$\pm$ 2.5    &  110    & 
410    & 46.83    &  0.35    &  0.04    &$2.4\times 10^{  -2}$    &$1\times
10^{ 24}$       &$6\times 10^{ 29}$      \\
\hline \T
0.06\%                                  &$^{156}_{ 66}$Dy$ \to ^{156}_{
64}$Gd$^{**}$  & 1$^{- }$    &1946.375$\pm$0.006& -7.5$\pm$ 6.6$\pm$ 2.5    &  110
   &  211    & 50.24    &  7.93    &  0.34    &$2.6\times 10^{  -2}$   
&$8\times 10^{ 23}$       &$5\times 10^{ 29}$      \\ \T
                                        &                                      
    &             &                  &-14.0$\pm$ 6.6$\pm$ 2.5    &  110    & 
311    & 50.24    &  1.69    &  0.10    &$2.8\times 10^{  -2}$    &$2\times
10^{ 24}$       &$3\times 10^{ 30}$      \\ \T \B
                                        &                                      
    &             &                  &-15.5$\pm$ 6.6$\pm$ 2.5    &  110    & 
411    & 50.24    &  0.29    &  0.03    &$2.8\times 10^{  -2}$    &$5\times
10^{ 24}$       &$8\times 10^{ 30}$      \\
\hline \T
0.06\%                                  &$^{156}_{ 66}$Dy$ \to ^{156}_{
64}$Gd$^{**}$  & 0$^{- }$    &1952.385$\pm$0.007& -1.1$\pm$ 6.6$\pm$ 2.5    &  110
   &  210    & 50.24    &  8.38    &  0.35    &$2.6\times 10^{  -2}$   
&$2\times 10^{ 24}$       &$6\times 10^{ 29}$      \\ \T
                                        &                                      
    &             &                  & -1.5$\pm$ 6.6$\pm$ 2.5    &  110    & 
211    & 50.24    &  7.93    &  0.34    &$2.6\times 10^{  -2}$    &$4\times
10^{ 25}$       &$1\times 10^{ 31}$      \\ \T
                                        &                                      
    &             &                  & -7.8$\pm$ 6.6$\pm$ 2.5    &  110    & 
310    & 50.24    &  1.88    &  0.11    &$3.5\times 10^{  -2}$    &$9\times
10^{ 24}$       &$3\times 10^{ 30}$      \\ \T
                                        &                                      
    &             &                  & -8.0$\pm$ 6.6$\pm$ 2.5    &  110    & 
311    & 50.24    &  1.69    &  0.10    &$2.8\times 10^{  -2}$    &$1\times
10^{ 26}$       &$7\times 10^{ 31}$      \\ \T \B
                                        &                                      
    &             &                  & -9.4$\pm$ 6.6$\pm$ 2.5    &  110    & 
410    & 50.24    &  0.38    &  0.04    &$2.7\times 10^{  -2}$    &$2\times
10^{ 25}$       &$1\times 10^{ 31}$      \\
\hline \T
0.06\%                                  &$^{156}_{ 66}$Dy$ \to ^{156}_{
64}$Gd$^{**}$  & 1$^{- }$    &1962.037$\pm$0.012&  8.1$\pm$ 6.6$\pm$ 2.5    &  110
   &  211    & 50.24    &  7.93    &  0.34    &$2.6\times 10^{  -2}$   
&$8\times 10^{ 23}$       &$5\times 10^{ 29}$      \\ \T
                                        &                                      
    &             &                  &  1.6$\pm$ 6.6$\pm$ 2.5    &  110    & 
311    & 50.24    &  1.69    &  0.10    &$2.8\times 10^{  -2}$    &$2\times
10^{ 24}$       &$6\times 10^{ 29}$      \\ \T \B
                                        &                                      
    &             &                  &  0.2$\pm$ 6.6$\pm$ 2.5    &  110    & 
411    & 50.24    &  0.29    &  0.03    &$2.8\times 10^{  -2}$    &$5\times
10^{ 24}$       &$1\times 10^{ 30}$      \\ 
\hline \T
0.06\%                                  &$^{156}_{ 66}$Dy$ \to ^{156}_{
64}$Gd$^{**}$  & 1$^{+ }$    &1965.950$\pm$0.004& 12.4$\pm$ 6.6$\pm$ 2.5    &  110
   &  210    & 50.24    &  8.38    &  0.26    &$2.6\times 10^{  -2}$   
&$4\times 10^{ 22}$       &$4\times 10^{ 28}$      \\ \T
                                        &                                      
    &             &                  & 12.0$\pm$ 6.6$\pm$ 2.5    &  110    & 
211    & 50.24    &  7.93    &  0.34    &$2.6\times 10^{  -2}$    &$8\times
10^{ 23}$       &$8\times 10^{ 29}$      \\ \T
                                        &                                      
    &             &                  &  5.7$\pm$ 6.6$\pm$ 2.5    &  110    & 
310    & 50.24    &  1.88    &  0.09    &$3.5\times 10^{  -2}$    &$2\times
10^{ 23}$       &$4\times 10^{ 28}$      \\ \T
                                        &                                      
    &             &                  &  5.6$\pm$ 6.6$\pm$ 2.5    &  110    & 
311    & 50.24    &  1.69    &  0.10    &$2.8\times 10^{  -2}$    &$2\times
10^{ 24}$       &$9\times 10^{ 29}$      \\ \T \B
                                        &                                      
    &             &                  &  4.2$\pm$ 6.6$\pm$ 2.5    &  110    & 
410    & 50.24    &  0.38    &  0.03    &$2.7\times 10^{  -2}$    &$3\times
10^{ 23}$       &$1\times 10^{ 29}$      \\
\hline
\T 0.06\%                                  &$^{156}_{ 66}$Dy$ \to ^{156}_{
64}$Gd$^{**}$  &  [$0^+$]    &1970.2 $\pm$0.8   & 16.7$\pm$ 6.6$\pm$ 2.5    &  110
   &  210    & 50.24    &  8.38    &  0.35    &$2.6\times 10^{  -2}$   
&$1\times 10^{ 23}$       &$2\times 10^{ 29}$      \\ \T
                                        &                                      
    &             &                  & 16.3$\pm$ 6.6$\pm$ 2.5    &  110    & 
211    & 50.24    &  7.93    &  0.34    &$2.6\times 10^{  -2}$    &$1\times
10^{ 25}$       &$2\times 10^{ 31}$      \\ \T
                                        &                                      
    &             &                  & 10.0$\pm$ 6.6$\pm$ 2.5    &  110    & 
310    & 50.24    &  1.88    &  0.11    &$3.5\times 10^{  -2}$    &$6\times
10^{ 23}$       &$3\times 10^{ 29}$      \\ \T
                                        &                                      
    &             &                  &  9.8$\pm$ 6.6$\pm$ 2.5    &  110    & 
311    & 50.24    &  1.69    &  0.10    &$2.8\times 10^{  -2}$    &$4\times
10^{ 25}$       &$3\times 10^{ 31}$      \\ \T \B
                                        &                                      
    &             &                  &  8.4$\pm$ 6.6$\pm$ 2.5    &  110    & 
410    & 50.24    &  0.38    &  0.04    &$2.7\times 10^{  -2}$    &$1\times
10^{ 24}$       &$7\times 10^{ 29}$      \\
 \hline \hline
 \end{tabular}
 \end{table}

\end{landscape}

\begin{landscape}

\begin{table}[t]
\vspace{-4mm}
\caption{Continued from Table \ref{tab:table3}.}
\label{tab:table44} 
\centering
\renewcommand{\arraystretch}{1.1}
\addtolength{\tabcolsep}{1pt}
\scriptsize
\vspace{2mm}
\begin{tabular}{|c|c|c|c|r|c|c|r|r|r|r|c|c|c|c|c|c|}
\hline \hline
     NA  \T \B &   Transition  & $J^{\pi}_f$ & $M^*_{A,Z-2} - M_{A,Z - 2}$ &
$M^{**}_{A,Z-2} - M_{A,Z}$ & $(n 2j l)_{\alpha}$ & $(n 2j l)_{\beta}$ &
$\epsilon_{\alpha}^*\;\;$ & $\epsilon_{\beta}^*\;\;$ & $\epsilon_{C}$ &
$\Gamma_{\alpha \beta}\;\;\;\;\;\;$  & $\tilde{T}_{1/2}^{\min}$ &
$\tilde{T}_{1/2}^{\max}$ \\
\hline \hline
\T 0.06\%                                  &$^{156}_{ 66}$Dy$ \to ^{156}_{
64}$Gd$^{**}$  & 0$^{+ }$    &1988.5  $\pm$0.2  & -7.0$\pm$ 6.6$\pm$ 2.5    &  210
   &  210    &  8.38    &  8.38    &  0.21    &$7.6\times 10^{  -3}$   
&$5\times 10^{ 23}$       &$3\times 10^{ 30}$      \\ \T \B
                                        &                                      
    &             &                  & -7.4$\pm$ 6.6$\pm$ 2.5    &  210    & 
211    &  8.38    &  7.93    &  0.22    &$7.7\times 10^{  -3}$    &$2\times
10^{ 25}$       &$2\times 10^{ 32}$      \\
       \T                                 &                                      
    &             &                  &-13.6$\pm$ 6.6$\pm$ 2.5    &  210    & 
310    &  8.38    &  1.88    &  0.10    &$1.7\times 10^{  -2}$    &$2\times
10^{ 24}$       &$6\times 10^{ 30}$      \\ \T
                                        &                                      
    &             &                  &-13.8$\pm$ 6.6$\pm$ 2.5    &  210    & 
311    &  8.38    &  1.69    &  0.09    &$9.4\times 10^{  -3}$    &$8\times
10^{ 25}$       &$8\times 10^{ 32}$      \\ \T \B
                                        &                                      
    &             &                  &-15.1$\pm$ 6.6$\pm$ 2.5    &  210    & 
410    &  8.38    &  0.38    &  0.03    &$8.7\times 10^{  -3}$    &$2\times
10^{ 24}$       &$3\times 10^{ 31}$      \\
\hline \T
0.06\%                                  &$^{156}_{ 66}$Dy$ \to ^{156}_{
64}$Gd$^{**}$  & 1$^{+ }$    &2026.664$\pm$0.006& 17.9$\pm$ 6.6$\pm$ 2.5    &  310
   &  311    &  1.88    &  1.69    &  0.07    &$1.8\times 10^{  -2}$   
&$3\times 10^{ 25}$       &$1\times 10^{ 32}$      \\ \T
                                        &                                      
    &             &                  & 16.5$\pm$ 6.6$\pm$ 2.5    &  310    & 
410    &  1.88    &  0.38    &  0.02    &$1.8\times 10^{  -2}$    &$4\times
10^{ 24}$       &$2\times 10^{ 31}$      \\ \T
                                        &                                      
    &             &                  & 16.4$\pm$ 6.6$\pm$ 2.5    &  310    & 
411    &  1.88    &  0.29    &  0.03    &$1.8\times 10^{  -2}$    &$7\times
10^{ 25}$       &$3\times 10^{ 32}$      \\ \T
                                        &                                      
    &             &                  & 16.3$\pm$ 6.6$\pm$ 2.5    &  311    & 
410    &  1.69    &  0.38    &  0.03    &$1.0\times 10^{  -2}$    &$4\times
10^{ 25}$       &$5\times 10^{ 32}$      \\ \T \B
                                        &                                      
    &             &                  & 16.2$\pm$ 6.6$\pm$ 2.5    &  311    & 
411    &  1.69    &  0.29    &  0.02    &$1.1\times 10^{  -2}$    &$7\times
10^{ 26}$       &$8\times 10^{ 33}$      \\
\hline \T
1.56\%                                  &$^{164}_{ 68}$Er$ \to ^{164}_{ 66}$Dy$^{*}
 $  & 0$^{+ }$    &    0             & -5.5$\pm$ 3.1$\pm$ 2.5    &  210    & 
210    &  9.05    &  9.05    &  0.22    &$8.6\times 10^{  -3}$    &$4\times
10^{ 23}$       &$1\times 10^{ 30}$      \\ \T
                                        &                                      
    &             &                  & -5.9$\pm$ 3.1$\pm$ 2.5    &  210    & 
211    &  9.05    &  8.58    &  0.23    &$8.3\times 10^{  -3}$    &$2\times
10^{ 25}$       &$5\times 10^{ 31}$      \\ \T
                                        &                                      
    &             &                  &-12.6$\pm$ 3.1$\pm$ 2.5    &  210    & 
310    &  9.05    &  2.05    &  0.11    &$1.8\times 10^{  -2}$    &$1\times
10^{ 24}$       &$3\times 10^{ 30}$      \\ \T
                                        &                                      
    &             &                  &-12.8$\pm$ 3.1$\pm$ 2.5    &  210    & 
311    &  9.05    &  1.84    &  0.09    &$1.0\times 10^{  -2}$    &$6\times
10^{ 25}$       &$4\times 10^{ 32}$      \\ \T \B
                                        &                                      
    &             &                  & -6.4$\pm$ 3.1$\pm$ 2.5    &  211    & 
211    &  8.58    &  8.58    &  0.27    &$8.0\times 10^{  -3}$    &$1\times
10^{ 26}$       &$5\times 10^{ 32}$      \\
\hline \T
0.14\%                                  &$^{168}_{ 70}$Yb$ \to ^{168}_{
68}$Er$^{**}$  & 1$^{- }$    &1358.898$\pm$0.005&  3.9$\pm$ 4.4$\pm$ 2.5    &  110
   &  211    & 57.49    &  9.26    &  0.38    &$3.3\times 10^{  -2}$   
&$4\times 10^{ 23}$       &$7\times 10^{ 28}$      \\ \T
                                        &                                      
    &             &                  & -3.6$\pm$ 4.4$\pm$ 2.5    &  110    & 
311    & 57.49    &  2.01    &  0.11    &$3.5\times 10^{  -2}$    &$1\times
10^{ 24}$       &$2\times 10^{ 29}$      \\ \T \B
                                        &                                      
    &             &                  & -5.3$\pm$ 4.4$\pm$ 2.5    &  110    & 
411    & 57.49    &  0.37    &  0.04    &$3.4\times 10^{  -2}$    &$3\times
10^{ 24}$       &$5\times 10^{ 29}$      \\
\hline \T
0.14\%                                  &$^{168}_{ 70}$Yb$ \to ^{168}_{
68}$Er$^{**}$  & 0$^{+ }$    &1422.10 $\pm$0.03 & 12.0$\pm$ 4.4$\pm$ 2.5    &  210
   &  310    &  9.75    &  2.21    &  0.11    &$1.8\times 10^{  -2}$   
&$1\times 10^{ 24}$       &$2\times 10^{ 30}$      \\ \T
                                        &                                      
    &             &                  & 11.8$\pm$ 4.4$\pm$ 2.5    &  210    & 
311    &  9.75    &  2.01    &  0.10    &$1.1\times 10^{  -2}$    &$4\times
10^{ 25}$       &$2\times 10^{ 32}$      \\ \T
                                        &                                      
    &             &                  & 11.5$\pm$ 4.4$\pm$ 2.5    &  211    & 
310    &  9.26    &  2.21    &  0.10    &$1.8\times 10^{  -2}$    &$8\times
10^{ 25}$       &$2\times 10^{ 32}$      \\ \T
                                        &                                      
    &             &                  & 11.4$\pm$ 4.4$\pm$ 2.5    &  211    & 
311    &  9.26    &  2.01    &  0.13    &$1.1\times 10^{  -2}$    &$2\times
10^{ 26}$       &$8\times 10^{ 32}$      \\ \T \B
                                        &                                      
    &             &                  &  9.7$\pm$ 4.4$\pm$ 2.5    &  211    & 
410    &  9.26    &  0.45    &  0.04    &$1.0\times 10^{  -2}$    &$1\times
10^{ 26}$       &$5\times 10^{ 32}$      \\
\hline \T \B
0.13\%                                  &$^{180}_{ 74}$W$ \to ^{180}_{ 72}$Hf$^*
 $  & 0$^{+ }$    &         0        &-12.0$\pm$ 3.9$\pm$ 2.1    &  110    & 
110    & 65.35    & 65.35    &  1.26    &$7.2\times 10^{  -2}$    &$3\times
10^{ 22}$       &$4\times 10^{ 27}$      \\
\hline \T \B
0.02\%                                  &$^{184}_{ 76}$Os$ \to ^{184}_{ 74}$W$^{**}$  
& $(0)^{+ }$  &1322.152$\pm$0.022& 11.3$\pm$ 1.3$\pm$ 0.9    &  110    & 
110    & 69.53    & 69.53    &  1.31    &$8.0\times 10^{  -2}$    &$7\times
10^{ 26}$       &$2\times 10^{ 27}$      \\
\hline \T
0.014\%                                 &$^{190}_{ 78}$Pt$ \to ^{190}_{
76}$Os$^{**}$  &$(1,2,3)^+$  &1382.4$\pm$0.2    & 15.6$\pm$ 5.7$\pm$ 1.5    &  210
   &  310    & 12.97    &  3.05    &  0.13    &$2.2\times 10^{  -2}$   
&$3\times 10^{ 23}$       &$8\times 10^{ 29}$      \\ \T
                                        &                                      
    &             &                  & 15.4$\pm$ 5.7$\pm$ 1.5    &  210    & 
311    & 12.97    &  2.79    &  0.12    &$1.6\times 10^{  -2}$    &$1\times
10^{ 25}$       &$6\times 10^{ 31}$      \\ \T
                                        &                                      
    &             &                  & 13.2$\pm$ 5.7$\pm$ 1.5    &  210    & 
410    & 12.97    &  0.65    &  0.05    &$1.5\times 10^{  -2}$    &$5\times
10^{ 23}$       &$2\times 10^{ 30}$      \\ \T
                                        &                                      
    &             &                  &  3.2$\pm$ 5.7$\pm$ 1.5    &  310    & 
410    &  3.05    &  0.65    &  0.04    &$2.2\times 10^{  -2}$    &$3\times
10^{ 24}$       &$1\times 10^{ 30}$      \\ \T \B
                                        &                                      
    &             &                  &  0.8$\pm$ 5.7$\pm$ 1.5    &  410    & 
410    &  0.65    &  0.65    &  0.03    &$1.5\times 10^{  -2}$    &$9\times
10^{ 24}$       &$5\times 10^{ 30}$      \\
 \hline \hline
 \end{tabular}
 \end{table}

\end{landscape}

\noindent
The half-lives are further normalized to an effective neutrino mass of $|m_{\beta\beta}| = 1$ eV.
The normalized half-lives listed in
Tables~\ref{tab:table2},~\ref{tab:table3},~and~\ref{tab:table44} are then given
as
\begin{equation}
{\tilde T}_{1/2} = T_{1/2} \left|\frac{m_{\beta\beta} }{ 1^{\;}\mathrm{eV}
}\right|^2                          
\left|\frac{M^{0\nu}(J_f^{\pi})}{\mathcal{M}^{0\nu}(0_f^{+})}\right|^2.
\label{tildeT}
\end{equation}

The unitary limit, i.e. the minimum value of ${\tilde T}_{1/2}$ (denoted as
${\tilde T}^{\mathrm{min}}_{1/2}$) is given for a full mass degeneracy of 
initial and final atoms. The maximum value ${\tilde T}^{\mathrm{max}}_{1/2}$ is
obtained by substituting in Eq.~(\ref{G1}) the mass difference squared
\begin{equation}
\Delta M ^2 = (M_{A,Z - 2}^{**} - M_{A,Z})^2 + \Delta M_{\mathrm{expt}}^2,
\label{dm}
\end{equation}
where the last term accounts for the experimental errors of the masses and the nuclear excited state:
\begin{equation}
\Delta M_{\mathrm{expt}}^2 = \delta M_{A,Z - 2}^{2} + \delta M_{A,Z}^2 + \delta
E_{\rm nucl}^{x 2}. \label{ddm}
\end{equation}
The atomic mass differences are occasionally known with higher precision than
the  individual masses. The normalized half-lives, however, should stay
between the two bounds given by ${\tilde T}^{\mathrm{min}}_{1/2}$ and ${\tilde
T}^{\mathrm{max}}_{1/2}$.

There are some other $0\nu$ECEC transitions, which may deserve an extra
comment:

\begin{itemize}

\item {$^{ 78}_{ 36}$Kr$(0^+) \to ^{ 78}_{ 34}$Se$^{**}$}

    This transition has a decay ECEC Q-value of 2867.5 keV and appears in the
list of likely resonant transitions in Ref.~\cite{DERU}. The final nucleus has
an excited state at 2864 keV, however, with unknown spin and parity. The state,
however, $\gamma$ decays by 100\% into the $3^+$ state at 1054 keV state, which
essentially excludes a spin $J=0,1$ assignment for the 2864 keV state. This
transition is excluded from our list.

\item {$^{ 96}_{ 44}$Ru$(0^+) \to ^{ 96}_{ 42}$Mo$^{**}$}

    A mass degeneracy of initial and final atoms could occur for this
transition by considering the excitation of final nucleus with 2712.68~keV.
The angular momentum and parity of this excited state are not known yet.
However, it decays to state with $J^\pi = 5^+/6^+$, which excludes a spin
$J=0,1$ assignment for this state. Therefore, this transition is excluded from
the list as well.

\item {$^{ 106}_{ 48}$Cd$(0^+) \to ^{ 106}_{ 46}$Pd$^{**}$}

    This transition might be a resonant transition in the case the palladium
isotope remains in an excited nuclear state at 2717.59~keV. However, the state
$\gamma$ decays by 100\% into the $3^+$ state at 1557.68 keV state, which
again excludes a possibility of $J=0,1$ for this state.

\item {$^{112}_{ 50}$Sn$(0^+) \to ^{112}_{ 48}$Cd$^{**}(0^+)$}

    The transition through the excited level at 1871 keV ($0^+$) has been 
widely discussed in the literature~\cite{DERU}. As already noted earlier, the
masses and thereby the Q-value for the ground-state ECEC decay have been
re-measured with high accuracy to $Q=1919.82$ keV \cite{RAKH09}. A double K-shell
excitation of the final atom requires about 55 keV, thereby leaving not enough
energy to excite the 1871 keV state. This transition is therefore  not
considered anymore in
Tables~\ref{tab:table2},~\ref{tab:table3},~and~\ref{tab:table44}.

\item {$^{130}_{ 56}$Ba$ \to ^{130}_{ 54}$Xe$^{**}$}

{
    The excitation level of 2544.43 $\pm$0.08 keV with unknown spin-parity
gives a quite low half-life assuming $J^{\pi} = 0^{\pm}$ and a double capture
of K electrons. If higher electron levels are involved,
the degeneracy gets broken. The excited $^{130}_{ 54}$Xe$^*$ level decays by
100 units into the $2^+$ state at 536 keV and by 10 units to the ground state.
The strong ground state mode excludes a $0^+$ assignment.
Possible assignments are $ J \geq 1$. Two K electrons have $J=0$, 
so such options are forbidden by the conservation of angular momentum.
}

    There is a possibility of a resonant enhancement of the $0\nu$ECEC of
$^{130}$Ba by considering a nuclear excitation of $^{130}$Xe at  2608.43 keV
or 2622.32 keV. The angular momentum and parity of these states are not
clarified yet. Unfortunately, a favored possibility of $0^+$ excited state is
excluded, because the $\gamma$-decay of these levels feed into a $4^+$ state.
These transitions have not been considered in
Tables~\ref{tab:table2}~-~\ref{tab:table44}.

\item {$^{162}_{ 68}$Er$(0^+) \to ^{162}_{ 66}$Dy$^{**}(1^+)$}

    The decay of $^{162}_{ 68}$Er (ECEC Q-value 1845 keV) with the excitation
of the $1745.72$ $(J^{\pi}_f = 1^{+ })$ keV state in $^{162}_{ 66}$Dy is a
good candidate as far as energy matching is concerned. A double K-shell capture
requires about 110 keV, however, the transition  to a $1^+$ final state is
strictly Pauli-forbidden.

\item {$^{184}_{ 76}$Os$ \to ^{184}_{ 74}$W$^{**}(0)^+$}

In the transition to the 1322 keV excited state of $^{184}_{ 74}$W, the mass difference 
exceeds three standard deviations. Accordingly, we do not expect 
a complete degeneracy. However, the minimum half-life estimated for three standard deviations in the direction of a smaller mass difference 
is less than $10^{27}$ years. This decay is included in the tables. 
Shown in place of the unitary limit is the minimum half-life found in this way.

\end{itemize}

All systems in Tables~\ref{tab:table2},~\ref{tab:table3},~and~\ref{tab:table44}
start out from stable parent nuclei.  We also investigated unstable and
radioactive candidate nuclei,
in particular those with rather long half-lives. However, none of them
was found to be of practical use. Even if those nuclei could be
produced
in reasonably large quantities, activity levels would be prohibitive.

\subsection{Likely resonant $0\nu$EPEP transitions}

The selection criteria have also been applied  to the neutrinoless double
electron production (EPEP), where the two electrons are placed into a bound
state
above the occupied electron shell of the final atom (see Eq. (\ref{EPEP})), and
where the simultaneous nuclear excitation provides the mass degeneracy. 
Clearly, this process is expected to be rather unlikely, 
as it requires that a $Q$-value be extremely fine-tuned to a nuclear excitation. 
If such a case existed, the two electrons would be placed into any
of the upper most non-occupied electron shells of the final atom, whereby the
width of the resonant decay would then be controlled by the width
of the excited nucleus. The atomic de-excitation width would be comparatively small, 
or even zero in the case the electrons occupied the atomic ground state.

As an instructive test, we analyze the $0\nu$EPEP transition of
$^{148}_{\,60}$Nd$(0^+)$ (isotopic abundance: 5.8\%) to the 1920.97 keV excited $0^+$ state
in $^{148}_{\,62}$Sm. In this case the difference of the atomic masses is
$M^{*}_{A,Z+2}  - M_{A,Z} = 7.2 \pm 2.8 \pm 2.4$ keV. The ground-state configuration
of the  $^{148}_{ 62}$Sm daughter atom is  $4f^6 6s^2$, whereas the one of
~$^{148}_{ 60}$Nd is $4f^4 6s^2$. The $0\nu$EPEP transition to the $4f^6 6s^2$ ground
state of $^{148}_{ 62}$Sm is disallowed by angular momentum, as it requires the
{production} of two electrons with $l=3$.
The most favored $0\nu$EPEP transition would be an atomic excitation of
$^{148}_{\,62}$Sm, whereby the two electrons were placed into the $7s$ shell.
The atomic width is small, and we estimate that the unitary limit assuming 
the nuclear half-life of 0.1 ps. The calculated shortest half-life 
would be above $10^{27}$ y.

\subsection{Data analysis of candidate transitions}

Tables~\ref{tab:table2},~\ref{tab:table3},~and~\ref{tab:table44} also give a list of decays in which the unitary limit for the half-life 
of the atom turns out to be low, because the low probability of capture of electrons from higher orbits is compensated by a low probability 
of de-excitation of the electron shell. The probability of finding degeneracy with an accuracy comparable to the de-excitation width, of course, 
decreases with the de-excitation width. The number of such transitions, however, is sufficiently large, so it makes sense to evaluate the chances 
of detecting degeneracy with improving the accuracy of the measurements, taking into account all the states. We restricted ourselves to the capture 
of electrons from the orbits with principal quantum number no higher than 4. Higher states may also participate in the decays, however, we have not considered them, 
because data on the de-excitation widths of electron shells with principal quantum number higher than 4 are not available.

The objective of the analysis of this subsection is to estimate a priori probability of finding in future experiments 
an atom with a half-life of less than some fixed value.
 
The analysis is made by considering all transitions listed in Tables~\ref{tab:table2},~\ref{tab:table3},~and~\ref{tab:table44}, 
as well as the transitions associated with the capture of electrons from higher 
orbits, that passed our filters, but were not included in Tables~\ref{tab:table2},~\ref{tab:table3},~and~\ref{tab:table44}, 
as long as we list not more than 5 transitions for each pair of the nuclei, associated with the electron capture from lowest 
orbits.

In the spirit of the Bayesian method, the mass difference 
between initial and final atoms is regarded as a random variable. We assume further that this 
value, $\eta$, is normally distributed around the current experimental value $M_{A,Z - 2}^{**} - M_{A,Z}$ 
with a dispersion determined by the current experimental error $\Delta M_{\mathrm{expt}}$. 
Future measurement of the mass difference between pairs of atoms with an accuracy of $\Delta \mu$ 
is consistent with the hypothesis of the degeneracy of atoms with a probability
\begin{equation}
w(\Delta \mu) = \int_{-\Delta \mu}^{\Delta \mu} \frac{d\eta}{\sqrt{2\pi } \Delta M_{\mathrm{expt}}} 
\exp ( - \frac{(\eta - M_{A,Z - 2}^{**} + M_{A,Z})^2}{2 \Delta M_{\mathrm{expt}}^2}).
\label{TOHG}
\end{equation}

Suppose a measurement gives $\eta = 0$ with the accuracy of $\Delta \mu \sim \Gamma_{\alpha\beta}$. 
For the normal distribution of $\eta$, in the likely resonant case $(M_{A,Z - 2}^{**} - M_{A,Z})^2 \sim \Delta M_{\mathrm{expt}}^2$, 
and for a small $\Gamma_{\alpha\beta}$, the probability 
of finding the complete degeneracy and thereby of getting the unitary limit equals 
$w(\Delta \mu) \sim \Gamma_{\alpha\beta}/\Delta M_{\mathrm{expt}}$, in agreement with naive expectations.

The half-live ${\tilde T}_{1/2}$ of a particular transition may be written as
\begin{equation}
{\tilde T}_{1/2} = {\tilde T}_{1/2}^{\min}\frac{\Delta M^2 +
\Gamma_{\alpha\beta}^2/4}{\Gamma_{\alpha\beta}^2/4}, \label{TOH}
\end{equation}
where $\Delta M$ is defined in Eq.~(\ref{dm}). To have a decay time less than ${\tilde T}_{1/2}$, 
it is sufficient to claim experimentally $\eta= 0$ with an accuracy better than
\begin{equation}
\Delta \mu ({\tilde T}_{1/2}) = \frac{\Gamma_{\alpha\beta}}{2} \left(\frac{{\tilde T}_{1/2}}{{\tilde T}_{1/2}^{\min}} - 1 \right)^{1/2}.
\label{TOH2}
\end{equation}
The value of $\Delta \mu ({\tilde T}_{1/2})$ is the maximum error, at which the decay time is shorter than ${\tilde T}_{1/2}$.

The number of transitions, $n­_T$, to be found with half-lives below ${\tilde T}_{1/2}$ can be estimated by summing up the Gaussian probabilities
\begin{eqnarray}
n­_T = \sum w(\Delta \mu ({\tilde T}_{1/2})) &\equiv& \sum\left(
  \frac{1}{2}\mathrm{erf}(\frac{M_{A,Z-2}^{**}-M_{A,Z}+\Delta
\mu(\tilde{T}_{1/2})}{\sqrt{2}\Delta M_{\mathrm{expt}}}) \right. \nonumber \\ 
&& \;\;\;\;\;\;\left. - \frac{1}{2} \mathrm{erf}(\frac{M_{A,Z-2}^{**}-M_{A,Z}-\Delta
\mu(\tilde{T}_{1/2})}{\sqrt{2}\Delta M_{\mathrm{expt}}})
\right).
\label{TOH3}
\end{eqnarray}
The sum runs over the transitions with ${\tilde T}_{1/2}^{\min} < {\tilde T}_{1/2}$, listed in Tables~\ref{tab:table2},~\ref{tab:table3},~and~\ref{tab:table44}.

\begin{figure} [htb!]
\begin{center}
\includegraphics[width = 0.62\textwidth]{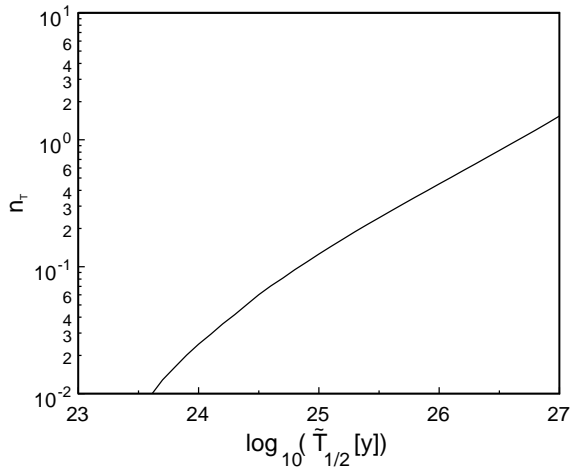}
\caption{ Expected number of transitions with a normalized half-life below ${\tilde
T}_{1/2}$, provided accurate measurements are made to determine the mass
difference between pairs of atoms listed in
Tables~\ref{tab:table2},~\ref{tab:table3},~and~\ref{tab:table44}. }
\label{fig:conf}
\end{center}
\end{figure}

The results are shown in Fig. \ref{fig:conf}. There are 2 chances out of 100 to find the transition with a normalized half-life below $10^{24}$ years, 
13 chances out of 100 to find the transition with a normalized half-life below $10^{25}$ years, 45 chances out of 100 to find the transition 
with a normalized half-life below $10^{26}$ years. We also expect $\sim 2$ transitions with ${\tilde T}_{1/2} < 10^{27}$ years.

Under the specified conditions, the chance of finding after precise mass measurements a $0\nu$ECEC decay more sensitive to the Majorana neutrino mass 
than the benchmark $^{76}$Ge neutrinoless $\beta ^-\beta^- $ decay is close to $\sim 1:25$. 
The experimental search for the $0\nu$ECEC decay is thus expected to be more complicated. 
The background conditions in the $0\nu$ECEC decay are, however, very favorable (see Sect. 4).

Here, all the transitions are treated as statistically independent. For most of the transitions in Tables~\ref{tab:table2},~\ref{tab:table3},~and~\ref{tab:table44}, 
this condition is certainly not satisfied, because for the same atom, 
together with the capture of electrons from the lower shells, one generally has captures of electrons from the high-lying levels. One precise measurement 
of the atomic mass difference either excludes the entire group, or allows all (or nearly all) transitions. 
Breit-Wigner resonances corresponding to the different electron states have a typical width of less than $\sim 10$ eV. In our case, such resonances do not overlap and 
are summed up additively. An increase in the decay probability is equivalent to registering an excess of events. 
The number of events calculated from Eq. (\ref{TOH3}) takes into account this effect. 
Although the list of transitions splits into the correlated groups, 
the statistically independent count of the expected decays is justified. 
While the correlations do not influence $n_T$, they obviously increase the dispersion of $n_T$.

The accuracy of 10 eV in the measurement of atomic masses
will be achievable in the near future \cite{penning1}.
The electron binding energy depends on the local physical and chemical
environment. An interesting question is whether and how to manage the atomic structure in
such a way as to artificially implement the degeneracy of the atoms and
create conditions for the resonant enhancement.

The effects of finite density and temperature on the profile of the atomic
spectral lines were studied theoretically and experimentally
starting from the 1930$^\prime$s. To give an idea of the magnitude of the effect, we
give some numbers.
In ionized atoms of He I, photon energy emitted in the transition between the
levels $3p$ and $2s$ at a density of $10^{18}$cm$^{-3}$ and temperature $T =
5000$ K
changes by about 10 meV (see, e.g., \cite{SOBE}, Sect. 39). In metals, when
the temperature varies from 100 K to room temperature, levels $1s$ in Li, ...,
$4p$ in Rb
get shifted by several tens of meV \cite{Riffe}. These values are 2-3 orders of
magnitude smaller than the width of electron holes in the atoms,
so the influence of the environment in such a case is not sufficient
and, apparently, is not of interest.

The chemical composition of matter also may influence the binding energies. In
comparison with the free atoms, the energy of atomic electrons in metals
gets shifted by a few eV \cite{Shirley}. The total effect on the mass
difference of atoms connected by double-electron capture can reach $\sim Z$ eV, 
which exceeds the typical natural width $\sim 10$ eV of
the electron holes in medium-heavy and heavy atoms. The effect
of chemical composition deserves more detailed analysis.

\section{Experimental signatures of $0\nu$ECEC decay}

Double beta (\dbb) decay of any variant is a process with a notoriously small
probability, where typical decay times are at least of the order $10^{19}$ y
in the most fortuitous cases. These values pertain to the ordinary cases, where
$2\nu$ are involved. For the more relevant neutrinoless decays,
half-lives are expected to be even longer by at least another 3 to 4  orders of
magnitude owing to the fact that the rate scales with the square of the mass
of the Majorana neutrino. To {detect} these extremely rare processes
in a real experiment, requires an enormous effort in background reduction.
There are two typical and rather different sources of  backgrounds one has to
deal with, cosmic ray interactions (including cosmogenically produced
radio-isotopes) and the ambient radioactivity. Even in underground laboratories
like the ones at Gran Sasso or at Modane, which provide a natural 3100 meter,
resp. 4800 meter water equivalent shielding against cosmic rays, an
experimental setup for \dbb\ decay measurements still requires a massive active
and passive shielding before any attempt can be made to venture into life-times
of the order of $10^{24}$ y and above~\cite{gerda,cuore,cobra}. Typical probe
masses are then still several tens or hundreds of kilograms.

Double beta decay experiments so far have focused almost entirely on the
$\beta^-\beta^-$-decay variant and little attention has been devoted to a
possible decay from the $\beta^+$ direction. Indeed, the 
$\beta^-\beta^-$-decay comes with a rather simple and comparatively easy to
detect signature: in the  $2\nu\beta^-\beta^-$-decay the summed  energy
spectrum of the two electrons carries the signature of a 4-body decay, whereas
a single, mono-energetic peak at the endpoint energy  signals the neutrinoless
decay. However, because the summed electron energy of the
$2\nu\beta^-\beta^-$-decay extends all the way to the endpoint region, any
background in this region becomes a serious issue. The spectral resolution of
the experiment is then most critical parameter, which determines the
sensitivity of the experiment.

The signatures of a decay from the $\beta^+$ direction are rather different. We
will not enter into a detailed discussion about  $\beta^+\beta^+$ or
$\beta^+$EC decay, as there is an extra energy penalty of $2m_ec^2$ to be paid
for each $\beta^+$ production on top of a rather disadvantageous $\beta^+$
phase-space factor {and final-state Coulomb repulsion}.

We focus on the ECEC process instead.
{In comparison with the $\beta^-\beta^-$-decay,
prospects for measuring the ECEC-decay looked at the beginning rather
pessimistic.}
Generally low isotopic abundances add to the difficulties. At present, there is
only one experiment, which has advanced quite significantly. This is the TGV
experiment~\cite{tgv} in the Modane underground laboratory looking for the ECEC
decay of $^{106}$Cd through the identification of $\approx 20$ keV X-rays. The
collaboration has already reached an impressive lower limit for the $2\nu$ECEC
decay of $T_{1/2}>2.6\cdot10^{20}$~y (90\%). On the other hand, the
neutrinoless  ECEC decay usually requires an extra photon, which must carry
away the excess energy given by the Q-value of the transition (which is about
2.7 MeV in the $^{106}$Cd case). This causes an extra suppression and the
expected half-life is above $10^{30}$ years for $m_{\beta\beta}=1$ eV by
referring to a calculation performed for $^{112}$Sn \cite{sujwy}.

The ECEC process discussed in this paper brings in a new aspect, which could be
rather advantageous as far as the experimental conditions are concerned. If
Nature provides us with a mass degeneracy, as discussed in this paper, the
decay proceeds through a narrow resonance and the rate increases dramatically
giving half-lives which are  competitive to the neutrinoless $\beta^-\beta^-$
decays. On top of this, there are several additional advantages. The
de-excitation of the final excited nucleus proceeds in most cases through a
cascade of easy  to detect $\gamma$ rays. A two- or even higher-fold
coincidence setup can  cut down any background rate right from the beginning,
thereby requiring significantly  less active or passive shielding. For instance
in the $^{136}$Ba case, the 2315~keV level would go through the cascade
$2315(0+)\rightarrow818(2^+)\rightarrow\rm{~g.s.(0^+)}$ ($2\gamma$ rays) and in
the $^{156}$Gd case, the 1952 keV level would de-excite as
$1952(0-)\rightarrow1242(1^-)\rightarrow89(2^+)\rightarrow\rm{~g.s.(0^+)}$
($3\gamma$ rays). A mere detection of these $\gamma$ rays would already signal
the ECEC decay beyond any doubt, as there are no other background processes
feeding those particular nuclear levels. Further,  the lepton number conserving
ECEC decay with two neutrinos,
\begin{eqnarray}
(A, Z)  + e_b^- + e_b^- \to (A, Z - 2)^{**} + \nu_e + \nu_e, \label{2nECEC}
\end{eqnarray}
is strongly suppressed due to the almost vanishing phase space. Note that the
$2\nu$ phase space has a dependence of $Q^5$, with Q being the remaining excess
energy. For a Q-value in the low MeV region the half-life of the $2\nu$ECEC  is
about $10^{22}$ years \cite{domin}. For the considered ECEC transitions in
(\ref{2nECEC}) the Q-value is assumed to be  below a keV, which  already gives
a suppression by  at least another 15 orders of magnitude. Thus, the signature
for a $0\nu$ECEC resonant transition would not in any way be contaminated with
$2\nu$ decays. In fact, the detection of  a $\gamma$ ray de-excitation cascade 
provides an unambiguous signature for the neutrinoless decay.

\section{Conclusion}

The phenomenon of mixing and oscillations of atoms due to weak interaction with
the violation of the total lepton number and parity was investigated.
We can expect that the oscillations can be observed when the energy
difference between two atomic states is small.
The resonant situation can occur, if the daughter atom is in an excited atomic
or/and nuclear state.
The de-excitations
of the final state would then proceed through  the emission of X-rays (atomic
structure)
and/or $\gamma$-rays (nuclear structure), both of which could be used as an
experimental signature for the process. The theoretical framework,
which has been employed to describe these processes, is similar to that of
other oscillation processes, like the oscillations of neutrinos, neutral kaons
and $B$-mesons, or neutrons/antineutrons.

The concept was applied to the $0\nu$ECEC process where -- because of the final
state excitation --  mother and daughter atoms could be degenerate in mass.
An experimental detection of such a process would immediately prove the 
Majorana nature of neutrinos.
The process can, therefore, be regarded as an alternative to  the
$0\nu\beta^-\beta^-$ decay.
A list of likely resonant transitions was provided.
It was argued that accurate mass difference measurements are of paramount
importance in order to narrow down the possibilities.
In fact, the precision of masses, presently at a several keV level, must  be
improved by at least  one or  two  orders of magnitude.
This is certainly possible given the progress made over the last years in the
use of ion traps.
Some examples were quoted, where mass measurements have reached accuracies
$\sim 10$ eV.

In an ECEC process the daughter atom is left with two electron vacancies, but
remains otherwise neutral.
This extra atomic excitation energy was calculated using a relativistic
description of the  electron wave function,
where the Coulomb interaction of the two holes was taken into account. In the
relativistic description,
parity-violating transitions from mother to daughter become possible, such as
e.g., capture of an $s_{1/2}$ and
a $p_{1/2}$ electron leading to the $0^+\rightarrow0^+$ transition.
The electromagnetic and Auger decay widths of the atomic states were included
in the calculations. In the case of mass degeneracy,
the decay width determines the rate of transition, and the maximum rate
corresponds to the unitarity limit.

The selection rules for electron capture from high orbits were obtained. We
have shown that nuclear transitions with a change in the nuclear spin
$\Delta J \geq 2 $ are strongly suppressed. The capture of K electrons is most
likely process due to the large overlap
of electron wave function with the nucleus. At small distances the wave
functions of $ s_{1 / 2} $ and $ p_{1 / 2} $
electrons have similar asymptotes, so that the capture of L$_{2}$
electrons appears to be a feasible process comparable with the K and L$_1$ captures.
High shells are suppressed by a small value of the wave functions in the
nucleus, however, these states have smaller decay widths,
which compensates partly for the smallness of the unitary limit of the decay
rate of the initial atoms.

Explicit formulas for the relativistic matrix elements of the $0\nu$ECEC were given.

We analyzed the database of the nuclei and their excited states and made a list
of the most promising candidates for a resonant $ 0 \nu $ECEC transition. Assuming that $ | m_{\beta
\beta} | = 1$~eV, we arrived at the lower bound of the normalized half-lives $\sim 10^{22} $ years, which is
about a factor of $ 100$ or $ 1000 $
below predictions for a time of $ 0 \nu \beta^-\beta^-$ decay. Because of the uncertainty in the masses,
the range of allowed half-lives is broad, and reaches several orders of magnitude.
Precise measurements of the ground state masses, as well as additional
spectroscopic information on the excited states
(energy, spin and parity), are highly warranted to improve predictions of half-lives.

Our statistical estimates based on the data in
Tables~\ref{tab:table2},~\ref{tab:table3},~and~\ref{tab:table44} show that with
the improvement of mass measurements and
$ | m_{\beta \beta} | = 1$~eV there are about 10 chances out of 100 to find
transition with a normalized half-life below $10 ^{25}$ years and
50 chances out of 100 to find transition with a normalized half-life below $10 ^{26}$
years. One can expect two transitions with a normalized half-life below $10 ^{27}$ years.

Inverse transitions leading to the neutrinoless production of two bound
electrons also were considered in cases of a mass degeneracy. The transitions
are found to be suppressed, because the electrons would have to occupy high orbits, 
where the overlap with the nucleus is small.

The detection technique for identifying a $0\nu$ECEC-decay is rather different
from the  $0\nu\beta^-\beta^-$-decay, as there is no inherent background from
the $2\nu$-decay, which    existing experiments have to cope with.  Further, 
by exploiting the coincidence technique, in particular if the de-excitation of
the nucleus proceeds through a $\gamma$-ray cascade,  a significantly improved 
signal to  background ratio could be obtained, which would alleviate some of
the demands on a low-background facility.

After submission of the manuscript, new experimental results on precision measurement of
$Q$ values and new constraints on the $0\nu$ECEC half-lives have been reported 
\cite{new43,new44,new45,new46,new47}.

\section{Acknowledgments}
This work was supported in part by the DFG projects 436 SLK 17/298, RUS
113/721/0-3, and FR601/3-1,
by the Transregio Project TR27 "Neutrinos and Beyond" and the Graduiertenkolleg
GRK683,
by the VEGA Grant agency under the contract
No.~1/0249/03, by the grant of Scientific Schools of Russian Federation No.
4568.2008.2 and the RFBR project No. 09-02-91341.

\begin{appendix}

\section{Electron wave function inside nucleus}
\setcounter{equation}{0}

The relativistic wave function of electron in the Coulomb potential has the
form
\begin{equation}
\Psi _{\alpha m_\alpha}(\mathbf{x})=\left(
\begin{array}{l}
f_{\alpha }(r)\Omega _{jlm_\alpha}(\mathbf{n}) \\
(-)^{(1 + l - l^{\prime})/2}g_{\alpha ^{\prime }}(r)\Omega _{jl^{\prime }
m_\alpha}(\mathbf{n})
\end{array}
\right),  \label{DIWF}
\end{equation}
where $\alpha =(njl)$, $\alpha ^{\prime }=(njl^{\prime }),$ $l^{\prime}=2j-l,$
and $\Omega _{jlm}(\mathbf{n})$ are spherical spinors.

In the relativistic Coulomb problem, one can distinguish three different
scales: The first scale is related to the Bohr radius $a_{B} = 1/(\alpha m Z)$.
This scale determines the normalization of the electron wave function. The
second scale $ 1/m $ is the Compton wavelength of an electron. Starting from
these distances down to zero the upper component of Dirac wave function differs
markedly from the non-relativistic Coulomb wave function. The size of the
nucleus is about 500 times smaller than the Compton wavelength of an electron,
so the effects associated with the finite size of the nucleus must be
calculated on the basis of the relativistic Dirac equation. One may select an
even smaller (third) scale. The radius of the nucleus is a few fermi, or
turning to the proton, its radius is about $1$ fm $\approx $ 1/200 MeV$^{-1}$,
which is less than $\alpha /m\approx 1/70$ MeV$^{-1}$. If $Z$ increases, the
radius of the nuclei $R=1.2 A^{1/3}$ fm increases as well, but $\alpha Z/m$ is
growing faster, so that $R < \alpha Z/m$ holds for all nuclei. The combination
of $\alpha Z/m$ defines a third scale below which the Coulomb potential is
higher than electron mass.

Asymptotics of the wave functions of electrons work for $r \lesssim
1/(2\lambda)$. Note that $\lambda \sim 1/(n a_B)$ in the non-relativistic case,
and $\lambda \sim m$ in the relativistic case. The nuclear radius is below
$1/(2\lambda)$ and even less than $\alpha Z/(2\lambda)$. The low $r$
approximation should therefore work very well.

In order to get the electron wave function inside the nucleus we use the
asymptotic expansion of the Dirac wave function at $2 \lambda r \ll 1$:
\begin{eqnarray}
f_{njl}(r) &\simeq& \frac{ \sqrt{2} \lambda ^{3/2} }{\Gamma(2\gamma + 1)}
(\alpha Z m/\lambda - \kappa - n_{r}) \sqrt{\frac{(m + \varepsilon) \Gamma
(2\gamma + n_{r} + 1) \lambda}{ \alpha Z m^2 ( \alpha Z m/ \lambda - \kappa)
\Gamma (n_{r} + 1)}}
(2\lambda r)^{\gamma - 1}, \nonumber \\
g_{njl}(r) &\simeq & - \frac{ \sqrt{2} \lambda ^{3/2} }{\Gamma(2\gamma + 1)}
(\alpha Z m/\lambda - \kappa + n_{r}) \sqrt{\frac{(m - \varepsilon) \Gamma
(2\gamma + n_{r} + 1) \lambda}{ \alpha Z m^2( \alpha Z m/ \lambda - \kappa)
\Gamma (n_{r} + 1)}} (2\lambda r)^{\gamma - 1}. \nonumber
\end{eqnarray}
The normalization and other conventions are those of Ref. \cite{BERE}. In
particular, $\gamma = \sqrt{(j + 1/2)^2 - (\alpha Z)^2}$, $\lambda = \sqrt{m^2
- \varepsilon^2}  $. The radial functions are real, since $\alpha Z m -\kappa
\lambda >0$. Inside the nucleus, the screening of the charge by the electrons
is weak, so $Z$ is the charge of the unscreened nucleus.

\begin{table}
\scriptsize
\centering
\addtolength{\tabcolsep}{-4pt}
\caption{The upper and lower radial functions of the Dirac bi-spinors, averaged
over the volume of the nucleus (in keV$^{3/2}$).} \label{tab:table4}
\begin{center}
\begin{tabular}{c c r r r r r r r r}
\hline \hline
Shell                     && $^{78}$Se$\;\;\;\;$  & $^{106}$Pd$\;\;\;\;$ &
$^{112}$Cd$\;\;\;\;$ & $^{120}$Sn$\;\;\;\;$ & $^{124}$Te$\;\;\;\;$ &
$^{130}$Xe$\;\;\;\;$ & $^{152}$Sm$\;\;\;\;$ & $^{156}$Gd$\;\;\;\;$ \\
\hline
$1s_{1/2}$             &$<f>$&     3.45$\times 10^{3}$ &  6.22$\times 10^{3}$ &
 6.80$\times 10^{3}$ &  7.42$\times 10^{3}$ &  8.83$\times 10^{3}$ & 
1.09$\times 10^{4}$ &  1.23$\times 10^{4}$ &  1.33$\times 10^{4}$ \\
                          &$<g>$&    -4.34$\times 10^{2}$ & -1.07$\times
10^{3}$ & -1.23$\times 10^{3}$ & -1.40$\times 10^{3}$ & -1.81$\times 10^{3}$ &
-2.47$\times 10^{3}$ & -2.94$\times 10^{3}$ & -3.30$\times 10^{3}$ \\
$2s_{1/2}$          &$<f>$&     1.25$\times 10^{3}$ &  2.31$\times 10^{3}$ & 
2.54$\times 10^{3}$ &  2.79$\times 10^{3}$ &  3.35$\times 10^{3}$ & 
4.19$\times 10^{3}$ &  4.77$\times 10^{3}$ &  5.20$\times 10^{3}$ \\
                          &$<g>$&    -1.58$\times 10^{2}$ & -4.00$\times
10^{2}$ & -4.59$\times 10^{2}$ & -5.26$\times 10^{2}$ & -6.87$\times 10^{2}$ &
-9.48$\times 10^{2}$ & -1.14$\times 10^{3}$ & -1.29$\times 10^{3}$ \\
$3s_{1/2}$          &$<f>$&     6.83$\times 10^{2}$ &  1.26$\times 10^{3}$ & 
1.39$\times 10^{3}$ &  1.52$\times 10^{3}$ &  1.83$\times 10^{3}$ & 
2.29$\times 10^{3}$ &  2.61$\times 10^{3}$ &  2.85$\times 10^{3}$ \\
                          &$<g>$&    -8.60$\times 10^{1}$ & -2.18$\times
10^{2}$ & -2.51$\times 10^{2}$ & -2.87$\times 10^{2}$ & -3.76$\times 10^{2}$ &
-5.18$\times 10^{2}$ & -6.23$\times 10^{2}$ & -7.05$\times 10^{2}$ \\
$4s_{1/2}$          &$<f>$&     4.43$\times 10^{2}$ &  8.19$\times 10^{2}$ & 
8.99$\times 10^{2}$ &  9.87$\times 10^{2}$ &  1.19$\times 10^{3}$ & 
1.48$\times 10^{3}$ &  1.69$\times 10^{3}$ &  1.84$\times 10^{3}$ \\
                          &$<g>$&    -5.58$\times 10^{1}$ & -1.41$\times
10^{2}$ & -1.63$\times 10^{2}$ & -1.86$\times 10^{2}$ & -2.43$\times 10^{2}$ &
-3.36$\times 10^{2}$ & -4.04$\times 10^{2}$ & -4.57$\times 10^{2}$ \\
$2p_{1/2}$       &$<f>$&    -1.72$\times 10^{1}$ & -6.00$\times 10^{1}$ &
-7.22$\times 10^{1}$ & -8.64$\times 10^{1}$ & -1.23$\times 10^{2}$ &
-1.87$\times 10^{2}$ & -2.37$\times 10^{2}$ & -2.78$\times 10^{2}$ \\
                          &$<g>$&    -1.37$\times 10^{2}$ & -3.47$\times
10^{2}$ & -3.99$\times 10^{2}$ & -4.57$\times 10^{2}$ & -5.97$\times 10^{2}$ &
-8.25$\times 10^{2}$ & -9.92$\times 10^{2}$ & -1.12$\times 10^{3}$ \\
$2p_{3/2}$      &$<f>$&     8.06$\times 10^{-1}$ &  2.07$\times 10^{0}$ & 
2.38$\times 10^{0}$ &  2.74$\times 10^{0}$ &  3.48$\times 10^{0}$ & 
4.62$\times 10^{0}$ &  5.66$\times 10^{0}$ &  6.31$\times 10^{0}$ \\
                          &$<g>$&    -5.02$\times 10^{-2}$ & -1.75$\times
10^{-1}$ & -2.10$\times 10^{-1}$ & -2.52$\times 10^{-1}$ & -3.46$\times
10^{-1}$ & -5.03$\times 10^{-1}$ & -6.49$\times 10^{-1}$ & -7.47$\times
10^{-1}$ \\
\hline \hline
\end{tabular}
\end{center}
\end{table}

For uniform distribution of nuclear density, the average values of the upper
and lower components of the electron wave functions inside the nucleus can 
easily be found:
\[
<f_{njl}(r)>=\frac{3}{\gamma +2}f_{njl}(R),
\]
\[
<g_{njl}(r)>=\frac{3}{\gamma +2}g_{njl}(R).
\]

At distances $r \lesssim \alpha Z/m$, the ratio (see, e.g., \cite{BERE})
\begin{equation}
\frac{f_{njl}(R)}{g_{njl^{\prime}}(R)}= \frac{\alpha Z}{\gamma + \kappa}
\approx \left\{
\begin{array}{ll}
-2\frac{l+1}{\alpha Z}, & j=l+1/2, \\
\frac{\alpha Z}{2l}, & j=l-1/2,
\end{array}
\right.
\end{equation}
is large in the $j=l+1/2$ states and is small in the $j=l-1/2$ states, provided
$\alpha Z \ll 1$. The electron wave functions with higher values of $j$ are
suppressed inside the nucleus by additional powers of $2R \lambda \ll 1$.

In the non-relativistic theory, wave functions of particles at small distances
behave like $\sim r^{l}$. Particles, however, become relativistic on the scale
of the order of the Compton wavelength. As a result, the short distance
behavior of the wave function changes. Instead of $\sim r^{l}$ we have $\sim
r^{\gamma - 1}$ ($\sim r^{j - 1/2}$ for $\alpha Z \ll 1$). The suppression of
the wave function at short distances is qualitatively different in
non-relativistic and relativistic theories.

Table \ref{tab:table4} shows the upper and lower radial components of the Dirac
bi-spinors, averaged over the volume of the nucleus.
The values are maximum for atoms with large $Z$ and for K electrons. One sees a
significant suppression of the wave functions of $2p_{3/2} $ electrons in
comparison with $ 1s_{1 / 2}$, $2p_{1 / 2} $, and $ 2s_{1 / 2} $ levels.

A more accurate calculation of the electron wave functions can be done on the basis of relativistic Dirac-Hartree-Fock approximation.

\section{Matrix elements for capture of $s_{1/2}$ and $p_{1/2}$ bound electrons}
\setcounter{equation}{0}

Because of the conservation of angular momentum, transitions from the state
$0^+$ to the excited states of nuclei with spin $J$ are possible only when the
captured atomic electrons have total angular momentum $J$. The upper and lower
components of Dirac bi-spinors behave at short distances as $\sim r^{\gamma -
1}$, where $\gamma = \sqrt{(j + 1/2)^2 - (\alpha Z)^2}$, $j$ is the total
angular momentum of electron. Wave functions of electrons with high angular
momentum are strongly suppressed inside the nucleus. Increase of $j$ by one
unit leads to suppression of the electron wave function inside the nucleus by
an amount $\sim R/a_B \ll 1$ where $R$ is the size of the nucleus and $a_B$ is
the Bohr radius. Thus, we restrict ourselves to estimates of matrix elements of
neutrinoless double capture of $s_{1/2}$ and $p_{1/2}$ electrons, whose wave
functions are given by Eq.~(\ref{DIWF}) with
\begin{eqnarray}
\Omega_{jlm}(\mathbf{n}) = \frac{1}{\sqrt{4 \pi}} \left\{ 
\begin{array}{ll}
\chi_{m}, &  j = 1/2,~l=0,\\ 
-i\mbox{\boldmath{$\sigma$}}\mathbf{n} \chi_{m}, & j = 1/2,~l=1,
\end{array}
\right. 
\end{eqnarray}
where $\chi_{m}$ are the Pauli spinors. We note that the $s_{1/2}$ wave function multiplied by $\gamma_5$
takes the form of the $p_{1/2}$ wave function if $g_{s_{1/2}}$ is replaced by $f_{p_{1/2}}$ 
and $f_{s_{1/2}}$ is replaced by $-g_{p_{1/2}}$.

By taking the advantage of the closure approximation the operator entering the
nuclear matrix element is of the form of the product of leptonic current and
two hadronic currents. We have
\begin{eqnarray}
&&  \sum_{m_\alpha m_\beta} C^{JM}_{j_\alpha m_\alpha~j_\beta m_\beta} ~ J_\mu
({\mathbf{x}}_1) J_\nu ({\mathbf{x}}_2) \left( {\Psi_{\alpha m_\alpha}}^T
({\mathbf{x}}_1) C \gamma^\mu \gamma^\nu (1 - \gamma_5 ) {\Psi_{\beta m_\beta}} ({\mathbf{x}}_2) \right. \nonumber \\
&-& \left. {\Psi_{\beta m_\beta}}^T ({\mathbf{x}}_1) C
\gamma^\mu \gamma^\nu (1 - \gamma_5 ) {\Psi_{\alpha m_\alpha}} ({\mathbf{x}}_2) \right)
= \left(1-(-1)^{j_\alpha+j_\beta - J}\right) \nonumber\\
&\times& \sum_{m_\alpha m_\beta} C^{JM}_{j_\alpha
m_\alpha~j_\beta m_\beta} ~{\Psi_{\alpha m_\alpha}}^T ({\mathbf{x}}_1) C
~\gamma^\mu \gamma^\nu~ (1 - \gamma_5 ) {\Psi_{\beta m_\beta}}
({\mathbf{x}}_2) ~ J_\mu ({\mathbf{x}}_1) J_\nu ({\mathbf{x}}_2).
\end{eqnarray}
Using the non-relativistic impulse approximation and exploiting the expansion
$\gamma_\mu\gamma_\nu = g_{\mu\nu} + i \sigma_{\nu\mu}$ we find
\begin{eqnarray}
&&\gamma_\mu J^\mu ({\mathbf{x}}_1) \gamma_\nu J^\nu ({\mathbf{x}}_2) = - g^2_A
\sum_n \sum_m \tau^-_n \tau^-_m  \delta({\mathbf{x}}_1 - {\mathbf{x}}_n)
\delta({\mathbf{x}}_2 - {\mathbf{x}}_m)
\times \nonumber\\
&&\left[ [-\frac{g^2_V}{g^2_A} + ({\mbox{\boldmath{$\sigma$}}}_n \cdot {\mbox{\boldmath{$\sigma$}}}_m) ]
\left(
\begin{array}{cc}
1 & 0 \\
0 & 1
\end{array}
\right)~+~ \frac{g_V} {g_A} \left(
\begin{array}{cc}
0 & \mbox{\boldmath{$\sigma$}}
\\
\mbox{\boldmath{$\sigma$}} & 0
\end{array}
\right) \cdot ({\mbox{\boldmath{$\sigma$}}}_n - {\mbox{\boldmath{$\sigma$}}}_m) \right. \nonumber \\ 
&&\left. + \left(
\begin{array}{cc}
i \mbox{\boldmath{$\sigma$}} &0 \\
0 & i \mbox{\boldmath{$\sigma$}}
\end{array}
\right) \cdot ({\mbox{\boldmath{$\sigma$}}}_n \times {\mbox{\boldmath{$\sigma$}}}_m) \right]. \label{twoJ}
\end{eqnarray}
In what follows, contributions of nuclear matrix elements associated with three
terms in Eq.~(\ref{twoJ})
are supplied with subscripts $S$ (scalar), $ tT $ (time component of the
tensor), and $ sT $ (the spatial component of the tensor), respectively.
The nuclear matrix elements of $ 0 \nu $ECEC are presented below.

\subsection{$0^+_i \rightarrow 0^+_f,1^+_f$ nuclear transitions}

The matrix element of the $0^+_i \rightarrow 0^+_f$ nuclear transition,
entering Eq.~(\ref{LNVP2}), splits into three parts
\begin{equation}
{\cal M}_{\alpha\beta}(0^+) = M^{(SP)}_{\alpha\beta}(0^+) +
M^{(tT)}_{\alpha\beta}(0^+) + M^{(sT)}_{\alpha\beta}(0^+).
\end{equation}
where
\begin{eqnarray}
M^{(SP)}_{\alpha\beta}(0^+) &=& <0^+_f \parallel \sum_{n~m} \tau^-_n \tau^-_m 
h(r_{nm}) \left[ F^{(+)}_{\alpha\beta}(r_n,r_m) +
G^{(+)}_{\alpha\beta}(r_n,r_m) ({\hat{\mathbf{r}}}_n\cdot{\hat{\mathbf{r}}}_m ) \right] \\
&\times&[-\frac{g^2_V}{ g^2_A} +({\mbox{\boldmath{$\sigma$}}}_n \cdot {\mbox{\boldmath{$\sigma$}}}_m) ]
\parallel 0^+_i>, \nonumber\\
M^{(tT)}_{\alpha\beta}(0^+) &=& i \frac{g_V}{g_A} <0^+_f \parallel \sum_{n~m}
\tau^-_n \tau^-_m  h(r_{nm}) G^{(+)}_{\alpha\beta}(r_n,r_m)
 ~(\mbox{\boldmath{$\sigma$}}_n-\mbox{\boldmath{$\sigma$}}_m)\cdot ({\hat{\mathbf{r}}}_n\times{\hat{\mathbf{r}}}_m )
\parallel 0^+_i>, \nonumber\\
M^{(sT)}_{\alpha\beta}(0^+) &=& <0^+_f \parallel \sum_{n~m} \tau^-_n \tau^-_m 
h(r_{nm}) ~G^{(+)}_{\alpha\beta}(r_n,r_m)
  ~(\mbox{\boldmath{$\sigma$}}_n\times \mbox{\boldmath{$\sigma$}}_m) \cdot ({\hat{\mathbf{r}}}_n\times{\hat{\mathbf{r}}}_m )
\parallel 0^+_i>. \nonumber
\end{eqnarray}
The neutrino exchange potential $h(r_{nm})$ is defined in Eq.~(\ref{npot}).

The transition matrix element into the $1^+$ state has three components
\begin{equation}
{\cal M}_{\alpha\beta}(1^+) = M^{(SP)}_{\alpha\beta}(1^+) +
M^{(tT)}_{\alpha\beta}(1^+) + M^{(sT)}_{\alpha\beta}(1^+),
\end{equation}
where
\begin{eqnarray}
M^{(SP)}_{\alpha\beta}(1^+) &=& <1^+_f \parallel \sum_{n~m} \tau^-_n \tau^-_m ~
h(r_{nm})~ G^{(-)}_{\alpha\beta}(r_n,r_m)~({\hat{\mathbf{r}}}_n \times {\hat{\mathbf{r}}}_m)  \\
&\times&[-\frac{g^2_V}{ g^2_A} +({\mbox{\boldmath{$\sigma$}}}_n \cdot {\mbox{\boldmath{$\sigma$}}}_m) ]
\parallel 0^+_i> \nonumber\\
M^{(tT)}_{\alpha\beta}(1^+) &=& i \frac{g_V}{g_A} <1^+_f \parallel \sum_{n~m}
\tau^-_n \tau^-_m  h(r_{nm}) [ F^{(-)}_{\alpha\beta}(r_n,r_m) 
(\mbox{\boldmath{$\sigma$}}_n - \mbox{\boldmath{$\sigma$}}_m) - G^{(-)}_{\alpha\beta}(r_n,r_m) \nonumber\\
&\times&\ [ {\hat{\mathbf{r}}}_n\cdot {\hat{\mathbf{r}}}_m (\mbox{\boldmath{$\sigma$}}_n - \mbox{\boldmath{$\sigma$}}_m) +
({\mbox{\boldmath{$\sigma$}}}_n -{\mbox{\boldmath{$\sigma$}}}_m )\cdot {\hat{\mathbf{r}}}_m ~{\hat{\mathbf{r}}}_n -
({\mbox{\boldmath{$\sigma$}}}_n -{\mbox{\boldmath{$\sigma$}}}_m)\cdot {\hat{\mathbf{r}}}_n ~{\hat{\mathbf{r}}}_m ]]
\parallel 0^+_i>\nonumber\\
M^{(sT)}_{\alpha\beta}(1^+) &=& <1^+_f \parallel \sum_{n~m} \tau^-_n \tau^-_m 
h(r_{nm}) [ F^{(-)}_{\alpha\beta}(r_n,r_m)~  \mbox{\boldmath{$\sigma$}}_n \times
\mbox{\boldmath{$\sigma$}}_m
 - G^{(-)}_{\alpha\beta}(r_n,r_m) \nonumber\\
&\times& [{\hat{\mathbf{r}}}_n\cdot {\hat{\mathbf{r}}}_m \mbox{\boldmath{$\sigma$}}_n \times
\mbox{\boldmath{$\sigma$}}_m - ({\mbox{\boldmath{$\sigma$}}}_n \times{\mbox{\boldmath{$\sigma$}}}_m)\cdot {\hat{\mathbf{r}}}_m
 {\hat{\mathbf{r}}}_n + ({\mbox{\boldmath{$\sigma$}}}_m \times{\mbox{\boldmath{$\sigma$}}}_n)\cdot {\hat{\mathbf{r}}}_n {\hat{\mathbf{r}}}_m ] ]
\parallel 0^+_i>. \nonumber
\end{eqnarray}

The functions $F^{(\pm)}_{\alpha\beta}(r_n,r_m)$ and
$G^{(\pm)}_{\alpha\beta}(r_n,r_m)$ depend on the quantum numbers of the
captured electrons
\begin{eqnarray*}
\frac{8\pi F_{\alpha \beta }^{(\pm)}(r_{n},r_{m})}{\sqrt{2}} &=&\left\{
\begin{array}{ll}
f_{\alpha }(r_{n})f_{\beta }(r_{m}) \pm f_{\beta }(r_{n}) f_{\alpha }(r_{m}), &
(n_{\alpha }s_{1/2},n_{\beta }s_{1/2}), \\
~~(f_{\alpha }(r_{n})g_{\beta }(r_{m}) \pm g_{\beta }(r_{n})f_{\alpha
}(r_{m}))/2 & \\
\pm (f_{\beta }(r_{n})g_{\alpha }(r_{m}) \pm g_{\alpha }(r_{n})f_{\beta
}(r_{m}))/2, &
(n_{\alpha }s_{1/2},n_{\beta }p_{1/2}), \\
g_{\alpha }(r_{n})g_{\beta }(r_{m}) \pm g_{\beta }(r_{n})g_{\alpha }(r_{m}), &
(n_{\alpha }p_{1/2},n_{\beta }p_{1/2}),
\end{array}
\right.  \\
\frac{8\pi G_{\alpha \beta }^{(\pm)}(r_{n},r_{m})}{\sqrt{2}} &=&\left\{
\begin{array}{ll}
g_{\alpha }(r_{n})g_{\beta }(r_{m}) \pm g_{\beta }(r_{n})g_{\alpha }(r_{m}), &
(n_{\alpha }s_{1/2},n_{\beta }s_{1/2}), \\
~~(-g_{\alpha }(r_{n})f_{\beta }(r_{m}) \mp f_{\beta }(r_{n})g_{\alpha
}(r_{m}))/2 & \\
\pm (-g_{\beta }(r_{n})f_{\alpha }(r_{m}) \mp f_{\alpha }(r_{n})g_{\beta
}(r_{m}))/2, &
(n_{\alpha }s_{1/2},n_{\beta }p_{1/2}), \\
f_{\alpha }(r_{n})f_{\beta }(r_{m}) \pm f_{\beta }(r_{n})f_{\alpha }(r_{m}), &
(n_{\alpha }p_{1/2},n_{\beta }p_{1/2}).
\end{array}
\right. \label{FPM}
\end{eqnarray*}
If the electrons have the same quantum numbers $\alpha = \beta$, these
functions should be divided further by $\sqrt{2}$.

It is worthwhile to note that for $\alpha=\beta$ (e.g., capture of two K
electrons)
$F^{(-)}_{\alpha\beta}(r_n,r_m)$ and $G^{(-)}_{\alpha\beta}(r_n,r_m)$ vanish.

Averaging of these functions over the nucleus with the unit weight also gives a zero result. However, 
these functions are multiplied by the nuclear matrix elements that are antisymmetric under permutation 
of the arguments. The outcome, therefore, is different from zero, although the naive factorization 
produces a vanishing result. 

A reasonable estimate can be obtained on the basis of Cauchy's inequality. 
Let $a_{ij}$ and $b_{ij}$ be antisymmetric tensors. The upper limit of
the sum $\sum_{ij}a_{ij}b_{ij}$ can be evaluated from
\[
\sum_{ij}a_{ij}b_{ij}
\leq \left( \sum_{i j}a_{ij}^{2}\right) ^{1/2} \times \left( \sum_{i
j}b_{ij}^{2}\right) ^{1/2}.
\]

As an estimate of the matrix element we take the upper limit that splits into a product of the lepton and nuclear parts. 
Combinations of the electron wave functions entering the matrix elements of $0^+_i \to 0^{\pm}_f$ and $1^{\pm}_f$ transitions are given in Table~1.

\subsection{$0^+_i \rightarrow 0^-_f,1^-_f$ nuclear transitions}

The matrix element decomposes into a sum of two parts that have different
tensor structure:
\begin{equation}
{\cal M}_{\alpha\beta}(0^-) = M^{(tT)}_{\alpha\beta}(0^-) +
M^{(sT)}_{\alpha\beta}(0^-),
\end{equation}
where
\begin{eqnarray}
M^{(tT)}_{\alpha\beta}(0^-) &=& i \frac{g_V}{g_A} <0^-_f \parallel \sum_{n~m}
\tau^-_n \tau^-_m  h(r_{nm})  \\
&\times&\left[ \left(~H^{(+)}_{\alpha\beta}(r_n,r_m)
~{\hat{\mathbf{r}}}_n ~-~ H^{(+)}_{\alpha\beta}(r_m,r_n)) ~{\hat{\mathbf{r}}}_m\right) \cdot
(\mbox{\boldmath{$\sigma$}}_n-\mbox{\boldmath{$\sigma$}}_m) \right]
\parallel 0^+_i>, \nonumber\\
M^{(sT)}_{\alpha\beta}(0^-) &=& <0^-_f \parallel \sum_{n~m} \tau^-_n \tau^-_m 
h(r_{nm}) \nonumber \\
&\times& \left[ \left(~H^{(+)}_{\alpha\beta}(r_n,r_m) ~{\hat{\mathbf{r}}}_n ~-~
H^{(+)}_{\alpha\beta}(r_m,r_n)) ~{\hat{\mathbf{r}}}_m\right) \cdot
(\mbox{\boldmath{$\sigma$}}_n\times\mbox{\boldmath{$\sigma$}}_m) \right]
\parallel 0^+_i>. \nonumber
\end{eqnarray}

The transition matrix element of the vector type splits into a sum of three
parts:
\begin{equation}
{\cal M}_{\alpha\beta}(1^-) = M^{(SP)}_{\alpha\beta}(1^-) +
M^{(tT)}_{\alpha\beta}(1^-) + M^{(sT)}_{\alpha\beta}(1^-),
\end{equation}
where
\begin{eqnarray}
M^{(SP)}_{\alpha\beta}(1^-) &=& - <1^-_f \parallel \sum_{n~m} \tau^-_n \tau^-_m
 h(r_{nm}) \\
&\times&\left[ H^{(-)}_{\alpha\beta}(r_n,r_m) {\hat{\mathbf{r}}}_n -
H^{(-)}_{\alpha\beta}(r_m,r_n) {\hat{\mathbf{r}}}_m \right] [-\frac{g^2_V}{ g^2_A}
+({\mbox{\boldmath{$\sigma$}}}_n \cdot {\mbox{\boldmath{$\sigma$}}}_m) ]
\parallel 0^+_i>, \nonumber \\
M^{(tT)}_{\alpha\beta}(1^-) &=& i \frac{g_V}{g_A} <1^-_f \parallel \sum_{n~m}
\tau^-_n \tau^-_m  h(r_{nm}) \nonumber \\
&\times& \left[ H^{(-)}_{\alpha\beta}(r_n,r_m)
{\hat{\mathbf{r}}}_n  \times (\mbox{\boldmath{$\sigma$}}_n-\mbox{\boldmath{$\sigma$}}_m) -
H^{(-)}_{\alpha\beta}(r_m,r_n) {\hat{\mathbf{r}}}_m \times (\mbox{\boldmath{$\sigma$}}_m-\mbox{\boldmath{$\sigma$}}_n)
\right]
\parallel 0^+_i>, \nonumber\\
M^{(sT)}_{\alpha\beta}(1^-) &=& <1^-_f \parallel \sum_{n~m} \tau^-_n \tau^-_m 
h(r_{nm}) \nonumber \\
&\times&\left[ H^{(-)}_{\alpha\beta}(r_n,r_m) {\hat{\mathbf{r}}}_n  \times
(\mbox{\boldmath{$\sigma$}}_n\times\mbox{\boldmath{$\sigma$}}_m) - H^{(-)}_{\alpha\beta}(r_m,r_n) {\hat{\mathbf{r}}}_m
\times (\mbox{\boldmath{$\sigma$}}_m\times\mbox{\boldmath{$\sigma$}}_n) \right]
\parallel 0^+_i>. \nonumber
\end{eqnarray}

The functions $H^{(\pm)}_{\alpha\beta}(r_n,r_m)$ have the form
\begin{eqnarray*}
\frac{ 8\pi H^{(\pm)}_{\alpha\beta}(r_n,r_m)}{\sqrt{2}}  =  \left\{
\begin{array}{ll}
g_{\alpha}(r_n) f_{\beta}(r_m) \pm g_{\beta}(r_n) f_{\alpha}(r_m) , &
(n_{\alpha }s_{1/2},n_{\beta }s_{1/2}), \\
~~(g_{\alpha}(r_n) g_{\beta}(r_m) \mp f_{\beta}(r_n) f_{\alpha}(r_m))/2  & \\
\pm (g_{\beta}(r_n) g_{\alpha}(r_m) \mp f_{\alpha}(r_n) f_{\beta}(r_m))/2 , &
(n_{\alpha }s_{1/2},n_{\beta }p_{1/2}), \\
f_{\alpha}(r_n) g_{\beta}(r_m) \pm f_{\beta}(r_n) g_{\alpha}(r_m) , & 
(n_{\alpha }p_{1/2},n_{\beta }p_{1/2}).
\end{array}
\right. \label{HPM}
\end{eqnarray*}
If $\alpha = \beta$, then the functions should be divided by $\sqrt{2}$. For
$\alpha = \beta$ (e.g., capture of two electrons from the K-shell), the
averaged matrix element of $H^{(-)}_{\alpha\beta}(r_n,r_m)$ vanishes.

It is important to note that due to parity violation, the transitions into the
$0^-_f$ states with the capture of two $s_{1/2}$ electrons
are allowed, as well as the transitions into the $0^+_f$ states with the
capture of one $s_{1/2}$ and one $p_{1/2}$ electrons.

The functions $H^{(+)}_{\alpha\beta}(r_n,r_m)$ and
$H^{(-)}_{\alpha\beta}(r_n,r_m)$ are explicitly symmetric and antisymmetric
with
respect to the indices $n$ and $m$. They are multiplied by the nuclear part of
the operator, which has the same symmetry properties.
Factorization of the transition amplitude $0^+_i \to 0^-_i$ is possible, since
the combination of electron wave functions is averaged
over the nuclear volume with a unit weight. The nuclear matrix element equals
(\ref{ZEMI}).

In the transition $0^+_i \to 1^-_f$ we encounter the same problem as in the
transition $0^+_i \to 1^+_f$. The nuclear part of the operator
is antisymmetric, so the straightforward factorization is impossible. We again
restrict ourselves to estimate the upper limit on the matrix
element squared. Before the estimate we must, however, take into account that
only the antisymmetric part of
$H^{(-)}_{\alpha\beta}(r_n,r_m)$ contributes to the amplitude, therefore, the
value of
\[
(H^{(-)}_{\alpha\beta}(r_n,r_m) - H^{(-)}_{\alpha\beta}(r_m,r_n))^2/4
\]
must be averaged.

The leptonic parts of the matrix elements discussed above are shown in Table
\ref{tab:table1}.

\end{appendix}


\end{document}